\documentclass[preprint]{aastex}
\usepackage{lscape}
\usepackage{natbib}
\usepackage{color}
\begin{document}
\title{Wide, Cool and Ultracool Companions to Nearby Stars from Pan-STARRS\,1}
\shorttitle{}
\author{Niall R. Deacon,\altaffilmark{1,2 }\email{deacon@mpia.de}
Michael C. Liu,\altaffilmark{2,3}
Eugene A. Magnier,\altaffilmark{2}
Kimberly M. Aller,\altaffilmark{2}
William M.J. Best,\altaffilmark{2}
Trent Dupuy,\altaffilmark{4,5}
Brendan P. Bowler,\altaffilmark{6,7,8,2}
Andrew W. Mann, \altaffilmark{9}
Joshua A. Redstone,\altaffilmark{10}
William S. Burgett,\altaffilmark{2}
Kenneth C. Chambers,\altaffilmark{2}
Peter W. Draper,\altaffilmark{11}
H. Flewelling,\altaffilmark{2}
Klaus W. Hodapp,\altaffilmark{12}
Nick Kaiser,\altaffilmark{2}
Rolf-Peter Kudritzki,\altaffilmark{2}
Jeff S. Morgan,\altaffilmark{2}
Nigel Metcalfe,\altaffilmark{11}
Paul A. Price,\altaffilmark{13}
John L. Tonry,\altaffilmark{2}
Richard J. Wainscoat\altaffilmark{2}
\altaffiltext{1}{Max Planck Institute for Astronomy, Koenigstuhl 17, D-69117 Heidelberg, Germany}
\altaffiltext{2}{Institute for Astronomy, University of Hawai`i, 2680 Woodlawn Drive, Honolulu, HI 96822, USA}
\altaffiltext{3}{Visiting Astronomer at the Infrared Telescope Facility, which is operated by the University of Hawaii under Cooperative Agreement no. NNX-08AE38A with the National Aeronautics and Space Administration, Science Mission Directorate, Planetary Astronomy Program}
\altaffiltext{4}{Harvard-Smithsonian Center for Astrophysics, 60 Garden Street, Cambridge, MA 02138, USA}
\altaffiltext{5}{Hubble Fellow}
\altaffiltext{6}{California Institute of Technology, Division of Geological and Planetary Sciences, 1200 East California Blvd, Pasadena, CA 91125, USA}
\altaffiltext{7}{Visiting Astronomer, Kitt Peak National Observatory, National Optical Astronomy Observatory, which is operated by the Association of Universities for Research in Astronomy (AURA) under cooperative agreement with the National Science Foundation.}
\altaffiltext{8}{Caltech Joint Center for Planetary Astronomy Fellow}
\altaffiltext{9}{Harlan J. Smith Fellow, Department of Astronomy, The University of Texas at Austin, Austin, TX 78712, USA}
\altaffiltext{10}{Equatine Labs, 89 Antrim St., \#2, Cambridge, MA 02139, USA}
\altaffiltext{11}{Department of Physics, University of Durham, South Road, Durham DH1 3LE, UK}
\altaffiltext{12}{Institute for Astronomy, University of Hawai`i, 640 N. Aohoku Place, Hilo, HI 96720, USA}
\altaffiltext{13}{Princeton University Observatory, 4 Ivy Lane, Peyton Hall, Princeton University, Princeton, NJ 08544, USA}}
 \label{firstpage}
 \begin{abstract}
We present the discovery of 61 wide ($>$5\arcsec) separation, low-mass
(stellar and substellar) companions to stars in the solar neighborhood
identified from Pan-STARRS\,1 (PS1) data and the spectral classification of 27
previously known companions. Our companions represent a selective subsample of
promising candidates and span a range in spectral type of K7--L9 with the
addition of  one DA white dwarf. These were identified primarily from a dedicated common proper motion search around nearby stars, along with a few as serendipitous discoveries from our Pan-STARRS\,1 brown dwarf search. Our discoveries include 24 new L dwarf companions and one known L dwarf not previously identified as a companion. The primary stars around which we searched for companions come from a list of bright stars with well-measured parallaxes and large proper motions from the {\it Hipparcos} catalog (8583 stars, mostly A--K~dwarfs) and fainter stars from other proper motion catalogues (79170 stars, mostly M~dwarfs). We examine the likelihood that our companions are chance alignments between unrelated stars and conclude that this is unlikely for the majority of the objects that we have followed-up spectroscopically. We also examine the entire population of ultracool ($>$M7) dwarf companions and conclude that while some are loosely bound, most are unlikely to be disrupted over the course of $\sim$10~Gyr. Our search  increases the number of ultracool M dwarf companions wider than 300~AU  by 88\% and increases the number of L dwarf companions in the same separation range by 96\%. Finally, we resolve our new L dwarf companion to HIP~6407 into a tight (0.13\arcsec, 7.4\,AU) L1+T3 binary, making the system a hierarchical triple.  Our search for these key benchmarks against which brown dwarf and exoplanet atmosphere models are tested has yielded the largest number of discoveries to date.
\end{abstract}
 \keywords{stars: low-mass, brown dwarfs, surveys}
 \section{Introduction}
Wide ($\gtrsim$100~AU) binary companions have long been used as a tool for identifying and studying faint stellar and substellar objects. Such systems are relatively common, at least 4.4\% of solar-type stars having a companion wider than 2,000~AU \citep{Tokovinin2012} and $\sim$25\% having companions wider than 100\,AU \citep{Raghavan2010}. Indeed, the Sun's closest stellar neighbor Proxima Centauri is a $\sim$15,000~AU common proper motion companion to Alpha Centauri \citep{Innes1915}. These objects are an important population for understanding models of binary star formation. The widest systems may have been formed by capture within a young cluster \citep{Kouwenhoven2010}, a mechanism which has been used to explain the apparent increase in the number of companions per log separation bin at separations over 20,000~AU \citep{Dhital2010}. Another possibility is that these objects formed closer in and were pushed out to wider orbits by three-body interactions \citep{Delgado-Donate2004,Umbreit2005,Reipurth2012}. In this scenario the wide companion fraction should be higher for close binary systems. Indeed \cite{Law2010} have found that 45$^{+18}_{-16}$\% of wide M~dwarf systems were resolved as hierarchical triples with high resolution imaging. 

Wide binaries also provide test cases for characterizing stellar and substellar properties. As these systems likely formed from the same birth cluster, the companions will have the same metallicity and age as their host stars. Hence if one component has these parameters determined, the values can be applied to the other component. For example, wide M dwarf companions to FGK stars have been used as calibrators for spectroscopic determinations of M~dwarf metallicity relations \citep{Rojas-Ayala2010,Mann2013,Mann2014}. This ``benchmarking'' process is even more important for substellar objects as brown dwarfs lack a stable internal energy source and hence exhibit a degeneracy between their mass, luminosity and age. For substellar companions this degeneracy can be broken using the age of the primary (in combination with the bolometric luminosity derived from their absolute magnitude and spectrum of the secondary) to estimate the radius, mass and effective temperature of the secondary from evolutionary models. This effective temperature can then be compared to that derived from model fits to the secondary's spectrum, testing the agreement between atmospheric and evolutionary models ( e.g. \citealt{Saumon2007,Deacon2012}). There are also a handful of systems where the ultracool\footnote{A term typically used to mean objects of spectral type M7 or later. These may be free-floating planetary mass objects, brown dwarfs or very low-mass stars, depending on the spectral type and age of the object.} secondary itself is a binary. Such very rare systems are not simply ``age benchmarks" \citep{Liu2008}; the mass of the secondary being measured dynamically, providing the opportunity for even more stringent tests of theoretical models \citep{Dupuy2009b}.

As a result of their importance, wide substellar companions have been an active area of study in recent years. Many substellar companions have been identified as by-products of larger searches for brown dwarfs (e.g. \citealt{Burningham2009}) or by matching known brown dwarfs with catalogs of known stars (e.g. \citealt{Faherty2010,Dupuy2012}). Dedicated large-scale companion searches such as \cite{Pinfield2006} are more rare. A summary of discoveries prior to 2010 is presented in \cite{Faherty2010}. Since then wide-field surveys such as SDSS (\citealt{SDSS_DR9}; see studies by \citealt{Zhang2010,Dhital2010}) and WISE (\citealt{Wright2010}; see work by \citealt{Luhman2012,Wright2013}) have been used to identify wide companions to stars. As these surveys are either single epoch or taken over a short period of time, they often require additional datasets, such as 2MASS \citep{Skrutskie2006}, to allow the identification of companions from their common proper motion with the primary. Hence the ideal tool for identifying wide, low-mass companions to stars is a red-sensitive, wide-field, multi-epoch survey.

Pan-STARRS\,1 is a wide-field 1.8m telescope situated on Haleakala on Maui in the Hawaiian Islands. Run by a consortium of astronomical research institutions, it has been surveying the sky north of $\delta=-30^{\circ}$ since May 2010. Just over half of the telescope's operating time is reserved for the 3$\pi$ Survey, a multi-filter, multi-epoch survey of $\frac{3}{4}$ of the sky ($\sim$30,000\,sq.deg.). The Pan-STARRS\,1 photometric system is defined in \cite{Tonry2012} and consists of five filters used for the 3$\pi$ survey ($g_{P1}$, $r_{P1}$, $i_{P1}$, $z_{P1}$, and $y_{P1}$) as well as an extra-wide $w_{P1}$ filter specially designed for asteroid searches. It is the $y_{P1}$ filter, centered on 0.99~microns ($\delta \lambda = 70$\,nm) which makes Pan-STARRS\,1 ideal for surveying the local population of brown dwarfs. So far over 100 T~dwarfs have been identified (\citealt{Deacon2011}, \citealt{Liu2011}, Liu et al. in prep.), many in the early-T regime \citep{Best2013}. Such objects were often missed by previous surveys due to their indistinct colors in the near-infrared compared to background M dwarfs. However the addition of Pan-STARRS\,1 astrometry and far-red optical photometry disentangles them from the much larger number of M dwarfs with similar near-infrared colors. 

As a wide-field multi-epoch survey, Pan-STARRS\,1 also provides an ideal
dataset for identifying wide, common proper motion companions to nearby
stars. This is a natural extension of our search for nearby field brown dwarfs
 and involves searching the Pan-STARRS\,1 proper motion database for objects moving with a common proper motion to known stars. This approach has already led to the discovery of a wide T4.5 companion to
the nearby K dwarf HIP~38939 \citep{Deacon2012}. Additionally, objects
discovered in our field brown dwarf search can be cross-matched with catalogs
of nearby stars to serendipitously identify wide multiple systems. In
\cite{Deacon2012a}, we used this method to identify a wide T5 companion to the M dwarf LHS~2803 (simultaneously found by \citealt{Muzic2012}). We present here the results of our full search for wide cool and ultracool companions identified using PS1 proper motions.
\section{Identification in Pan-STARRS1 data}
\subsection{Primary star selection}
To identify objects with common proper motion to nearby stars, we first began by collating lists of stars to search around. We started with stars from the {\it Hipparcos} catalog \citep{vanLeeuwen2007}. To limit the contamination in our sample by distant background stars, we included only {\it Hipparcos} stars with proper motions above 0.1\arcsec/yr and parallaxes more significant than 5$\sigma$ (d$\lesssim$200\,pc). This provided us with a relatively complete sample of stars in the solar neighborhood with spectral types A--K. We supplemented this catalog with lower-mass primaries drawn from the LSPM (\citealt{Lepine2005}; $\mu >$0.15\arcsec/yr, $\delta >0^{\circ}$) and rNLTT (\citealt{Salim2003}; $\mu >$0.2\arcsec/yr) proper motion catalogs and with the bright M dwarfs catalog of \cite{Lepine2011} to which we applied a proper motion cut of $\mu >$0.1\arcsec/yr. See Table~\ref{surveys} for details on the number of primary stars from each input catalog.

\subsection{Selection of companions from Pan-STARRS\,1 data}
Pan-STARRS\,1 is an ongoing survey and as such the available data products are constantly evolving. We conducted our search over several iterations of the Pan-STARRS\,1~3$\pi$~database. In all our seaches we queried the most up-to-date Pan-STARRS\,1 3$\pi$ database using the {\it Desktop Virtual Observatory} software \citep{Magnier2008}. Initially, our search involved combining 2MASS and Pan-STARRS\,1 data to calculate proper motions in a similar process to that used by \cite{Deacon2011}. For this PS1 + 2MASS search, we required that objects had more than one detection in the $y_{P1}$ band. PS1 detections were required to have a significance greater than 5$\sigma$ to be included in the astrometric and photometric solutions of their parent object. We also required that the objects  be classified as good quality, point-source detections in both Pan-STARRS\,1 and 2MASS.

Since June 2012, we have used proper motions
calculated using Pan-STARRS\,1-only data (although 2MASS data is included if a
detection is within 1\arcsec~of the mean Pan-STARRS\,1 position). In these
cases, we required that the proper motion measurement of the candidate companion was
more significant than 5$\sigma$,  be calculated from more than seven
position measurements, and that the time-baseline over which it was calculated
 be greater than 400 days. Note that the Pan-STARRS\,1 survey strategy
consists of pairs of observations in the same filter taken $\sim$25~minutes
apart and often two filters will be taken in the same night. Hence our requirement
of more than seven position measurements does not imply more than seven independent epochs evenly spread
across the time-baseline of the proper motion calculation. 

In the case of the {\it Hipparcos} and \cite{Lepine2011} catalogs, distance estimates were available for all our target stars. Hence we searched for companions out to projected separations of 10,000~AU. For the LSPM and rNLTT catalogs, where distance estimates are not available for all the stars, we used a maximum search radius of 20 arcminutes. In all cases we made an initial cut on our candidates, requiring that candidates have proper motions that agree with their supposed primaries' proper motions to within 0.1\arcsec/yr in both R.A. and declination. This initial cut uses a tolerance that is much larger than our typical Pan-STARRS\,1 proper motion errors of $\sim$5 milliarcseconds per year. Next, we made much more stringent cuts on proper motion to select only likely companions. 

We restricted our search to objects which had proper motion differences compared to their primaries that were less significant than 5$\sigma$, here $\sigma$ is the quadrature sum of proper motion difference in each axis divided by the total proper motion error in that axis.

\begin{equation}
\label{sigmaequation}
\sigma^2 = \frac{(\mu_{\alpha,1}-\mu_{\alpha,2})^2}{\sigma_{\mu_{\alpha,1}}^2+\sigma_{\mu_{\alpha,2}}^2} + \frac{(\mu_{\delta,1}-\mu_{\delta,2})^2}{\sigma_{\mu_{\delta,1}}^2+\sigma_{\mu_{\delta,2}}^2}
\end{equation}
 Our typical proper motion errors in Pan-STARRS\,1 + 2MASS data are $\sim$5~millarcseconds per year. We selected two samples for follow-up observations. In all cases we set a zone
of exclusion around the Galactic Centre $|b|<$5$^{\circ}$ and $l<90$ or $l>270$. 
\begin{trivlist}
\item{{\bf 1. Ultracool companions:} Objects with red Pan-STARRS1 colors
($z_{P1}-y_{P1}>0.8$ mag) or ($y_{P1}-J_{2MASS}>1.8$ mag) were selected as candidate ultracool companions. These cuts
should only select objects of spectral types of M7 or later
\citep{Deacon2011}. These targets also had their Pan-STARRS\,1 and 2MASS
colors inspected to ensure they truly were red objects and had not entered
our sample due to erroneous $z_{P1}$ or $y_{P1}$ photometry. This involved removing objects with more than one $g_{P1}$ detection or with blue $i_{P1}-y_{P1}$ colours. 

Note that two objects (the companions to LSPM~J2153+1157 and NLTT~39312) are both close companions that did not meet our $z_{P1}-y_{P1}$ color cut but were selected due to their red $y_{P1}-J_{2MASS}$ colors ($>$1.6) and close proximity ($<12$\arcsec) to their primaries. These were subsequently spectrally typed as ultracool dwarfs. These two objects may have had their photometry affected by the proximity of their primary. 

This sample yielded a total of 36 new discoveries, one of which (HIP~13589B) was simultaneously discovered by \cite{Allen2012}. We have also reidentified and typed the NLTT~22073 system originally identified by \cite{Deacon2007} and HIP~78184~B originally proposed as a candidate companion by \cite{Pinfield2006}.}

\item{{\bf 2. Bright companions:} We selected the brightest of our  common proper motion candidates ($J_{2MASS}<15.5$ mag) for a poor-weather back-up program for our NASA IRTF SpeX
observations regardless of their color. These objects were mostly within an arcminute of the primary
with all having separations smaller than 3\arcmin. This resulted in the discovery of 20 new, bright companions and the typing of 24 previously known companions.}
\end{trivlist}

\subsection{Serendipitous discoveries}
As part of our ongoing search for T dwarfs in Pan-STARRS\,1 data (see
\citealt{Deacon2011} and \citealt{Liu2011} for more details), we have been obtaining follow-up
near-infrared photometry of Pan-STARRS\,1 selected candidates.
UKIDSS \citep{Lawrence2007} or VISTA \citep{Emerson2010} data were used where available; otherwise our candidate T dwarfs were observed using WFCAM
\citep{Casali2007} on the United Kingdom Infrared Telescope (UKIRT). We then
cross-matched all our observed candidates with our combined proper motion
catalogue from \cite{Lepine2005}, \cite{Salim2003}, and
\cite{Lepine2011} with a pairing radius of 20\arcmin. Objects that had
$y_{P1}-J_{2MASS}>2.2$~mag were initially selected to be T dwarf candidates. However if the additional near-IR photometry did not match our $y_{P1}-J_{MKO}>2.4$~mag and $J_{MKO}-H_{MKO}<0.7$~mag high priority T dwarf cut, but still met our $y_{P1}-J_{MKO}>2.0$~mag criterion, those objects were considered to be candidate L dwarfs. Any of these $\sim$500
candidates that then had proper motions  which did not deviate by more than 5$\sigma$ from their supposed
primary stars were selected as candidate L companions for this search.

 One of the serendipitous candidates we selected was a possible companion to the
late-type high proper motion star NLTT~730. After searching the Simbad
database, we identified our candidate object as 2MASS~J00150206+2959323, a
blue L7.5 dwarf found by \cite{Kirkpatrick2010} as a field object. As a blue L benchmark, this is similar to the G~203-50AB system identified by \cite{Radigan2008a} although it is substantially wider (5070\,AU vs. 135\,AU). This object is included in our subsequent analysis. In total we  selected five serendipitous candidate companions.

Our  proper motion companions are reported in Tables~\ref{all_info} and \ref{HIPsecondaries}. A
comparison between their proper motions and photometric distances (see
Section~\ref{phot_dist}) and the proper motions and distances of their
primaries are shown in Table~\ref{HIPsystems} for the {\it Hipparcos} companions and
Table~\ref{othersystems} for the companions to M~dwarfs. In all our tables, the proper motions from the secondaries come from Pan-STARRS\,1 + 2MASS proper motions. Where the object's 2MASS position was not included in the initial astrometric analysis we recalculated the proper motion solution including the 2MASS data. The proper motions and distances for all of our {\it Hipparcos} primaries come from \cite{vanLeeuwen2007}.

\section{Follow-up observations and archival data}
\subsection{Infrared photometry}
Where companions have confused or noisy photometry, we have used UKIRT/WFCAM \citep{Casali2007} to acquire additional near-infrared photometry. These
data were reduced at the Cambridge Astronomical Survey Unit using the WFCAM
survey pipeline \citep{Irwin2004, Hodgkin2009}. See
Table~\ref{HIPsecondaries_ir} for the near-infrared photometry of our
companions from our UKIRT observations, 2MASS \citep{Skrutskie2006}, UKIDSS \citep{Lawrence2007}, VISTA; \citep{Emerson2010}, pre-release photometry from the UKIRT Hemisphere Survey (UHS; Dye et al., in prep., accessed through the WFCAM Science Archive; \citealt{Hambly2008}) and Table~\ref{HIPsecondaries_wise} for mid-infrared photometry from WISE~\citep{Wright2010,WISE2012}.
\subsection{Near-infrared spectroscopy}
\label{IRspec}
To characterize our companions, we obtained 0.8--2.5~\micron\ spectroscopy
using the SpeX instrument \citep{Rayner2003} on the NASA Infrared Telescope
(IRTF). To minimize the possibility of observing an unrelated background
object due to a non-physical chance alignment with one of our primaries, we
preferentially followed up  (i.e. we were more likely to follow-up)
companions closer than five arcminutes  but did not follow-up all of
  our candidate companions. In total 115 candidate companions to stars were
passed to the IRTF queue, 85 are presented here as true companions, one we
classify as an unlikely companion and the remaining 29 were not observed.

Depending on the brightness of the object and the weather conditions, we used either the low-resolution ($R\approx$75--120) single-order prism mode or the moderate-resolution ($R\approx$750-2000) multiple-order cross-dispersed SXD mode. The slit width was chosen to match the seeing and was oriented along the parallactic angle to minimize atmospheric dispersion. The observations were taken in nodded ABBA patterns. The standards were taken contemporaneously with each science target and at similar airmass and sky position. We reduced all our spectra using version~3.4 of the SpeXtool software package \citep{Cushing2004, Vacca2003}. See Table~\ref{SpeXlog} for details of the exposure times, weather conditions and standard stars used for each object.

We spectrally classified our objects by visually comparing with the M and L
near-infrared standards from \cite{Kirkpatrick2010}. Additionally, for our L
dwarf companions we measured three flux indices (H$_2$O-J, H$_2$O-H and
CH$_4$-K) from \cite{Burgasser2006} relevant for L dwarfs. We then applied the
polynomial relations of \cite{Burgasser2007} to derive spectral types, averaging over the types derived from each index. We note
that the CH$_4$-K index does not depend strongly on spectral type earlier than
mid-L and hence excluded this from the averaging for objects with this visual classification. Figures~\ref{HIP_early_M_spectra}
and \ref{HIP_mid_M_spectra} show our early to mid-M~dwarf companions to
{\it Hipparcos} stars; Figure~\ref{HIP_ultracool_spectra} shows our late-M and L~dwarf {\it Hipparcos} companions; and Figure~\ref{other_spectra} shows our companions around non-{\it Hipparcos} primaries and serendipitous discoveries. Additionally, we observed a number of our primary stars that lacked spectral types in the literature. The spectra for these objects are shown in Figure~\ref{HIP_primaries}.

 For objects with spectral types earlier than M, there are no near-infrared standards so we compared with the spectral library of \cite{Cushing2005} and \cite{Rayner2009}, selecting the best comparison spectrum visually. For any of our objects with spectral types earlier than M5, we specifically examined the $\sim$0.85\,$\mu$m TiO feature in the $Z$-band and the $Y$-band 1\,$\mu$m FeH feature. For two primaries classified as G~dwarfs we also examined the strengths of metal and hydrogen lines. We identified one of our companions (HIP~88728~B) as having blue continua and weak Paschen lines. Hence we classify this object as a DA white dwarf.

\subsection{Optical spectroscopy}
\label{Optspec}
 As a number of our primaries were poorly characterised, we obtained optical spectroscopy from both the Kitt Peak Mayall 4-m telescope and the University of Hawaii 2.2-m telescope
\subsubsection{Kitt Peak Mayall 4-m spectroscopy}
 On 2013~Dec~31~UT we obtained optical spectra of NLTT~38389 and NLTT~1011 with the 
Ritchey-Chretien Spectrograph equipped with the T2KA CCD on the Kitt Peak National Observatory 
4-m Mayall telescope. The BL 420 grating was blazed at 7800\,\AA~(first order) with the GG495 
order blocking filter. The slit was set at 1$\farcs$5 by 98$''$, resulting in spectra spanning 
6300--8500~\AA at a resolving power of $\sim$3~\AA. NLTT~38389 and NLTT~1011 were targeted at airmasses of 1.25 and 1.01 with exposures
of 600 sec and 240 sec, respectively.  We also acquired optical spectroscopy for one of our companions (HIP~84840~B) on 2014~May~22~UT with the same set-up at airmass 1.03. The slit was oriented in the N-S direction for all
of the observations, so slit losses from chromatic dispersion may be non-negligible for NLTT~38389.
The spectrophotometric standards HR~3454 \citep{Hamuy1992} (2013~Dec~31~UT)  and HZ44 (2014~May~22~UT)  were observed at airmasses of 1.15 an 1.0 respectively for
flux calibration. Each 2D image was corrected for bad pixels and cosmic rays, bias-subtracted, and 
flat fielded. After sky subtraction, each spectrum was corrected for throughput losses using our 
standard star measurements.

\subsubsection{University of Hawaii 2.2m telescope/SNIFS spectroscopy}
 To aid in characterizing the primary stars, we took spectra of GD 280, NLTT 730, LSPM J2153+1157, and NLTT 22073 with the SuperNova Integral Field Spectrograph \citep{Aldering2002,Lantz2004} on the University of Hawaii 2.2-m telescope on Mauna Kea on UT 2013 October 19 and December 13-14. SNIFS provides simultaneous coverage from 3200\,\AA~to 9700\,\AA~at a resolution of $R\sim1000$. Integration times varied from 65 to 500\,s, depending on the $R$ magnitude. This was sufficient to get reasonable S/N ($>70$ at 6000\,\AA) on all targets except GD~280, which had particularly low S/N ($\sim15$) due to patchy cloud cover and a fainter magnitude. The SNIFS pipeline \citep{Bacon2001} performed basic reduction, including dark, bias, and flat-field corrections, wavelength calibration, sky subtraction, and extraction of the 1D spectrum. We took spectra of the  spectrophotometric standards EG~131, Fiege~110, GD~71, Feige~66, and HR~7596 during the night, which were then used to flux calibrate the data and remove telluric lines. Additional information on SNIFS data processing can be found in \citet{Lepine2013} and \citet{Mann2013a}.

\section{New common proper motion companions}
\begin{figure}[htbp]
\begin{center}
\epsscale{0.8}
\plotone{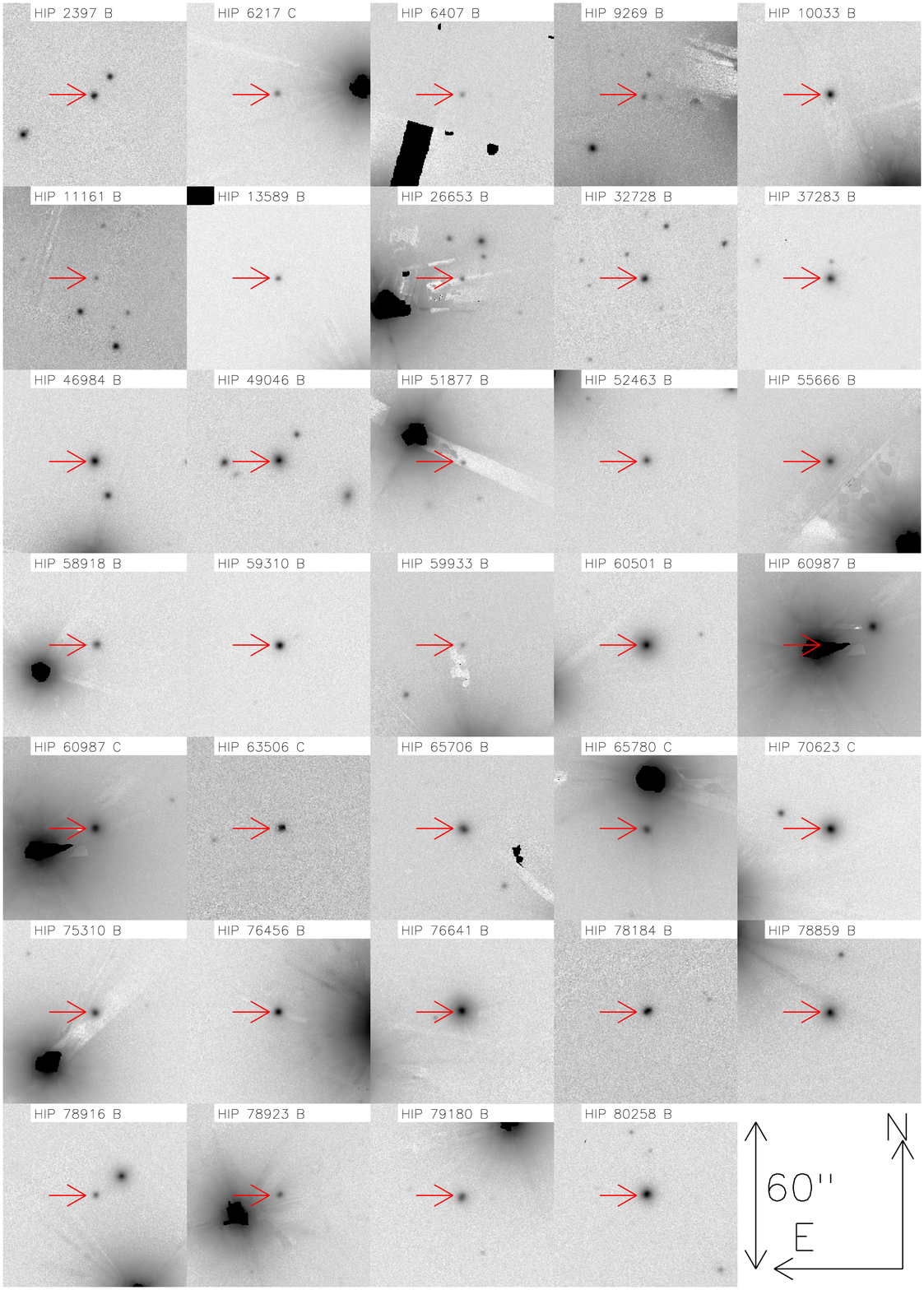}
\caption{\label{finders1} The $y_{P1}$ finder charts for our companions. Note that some regions are masked due to chip gaps and PS1 detector artifacts. Also note that some of the higher proper motion stars appear elongated, due to these images being stacks of individual observations spread over the PS1 survey period.}
\end{center}
\end{figure}

\begin{figure}[htbp]
\begin{center}
\epsscale{0.8}
\plotone{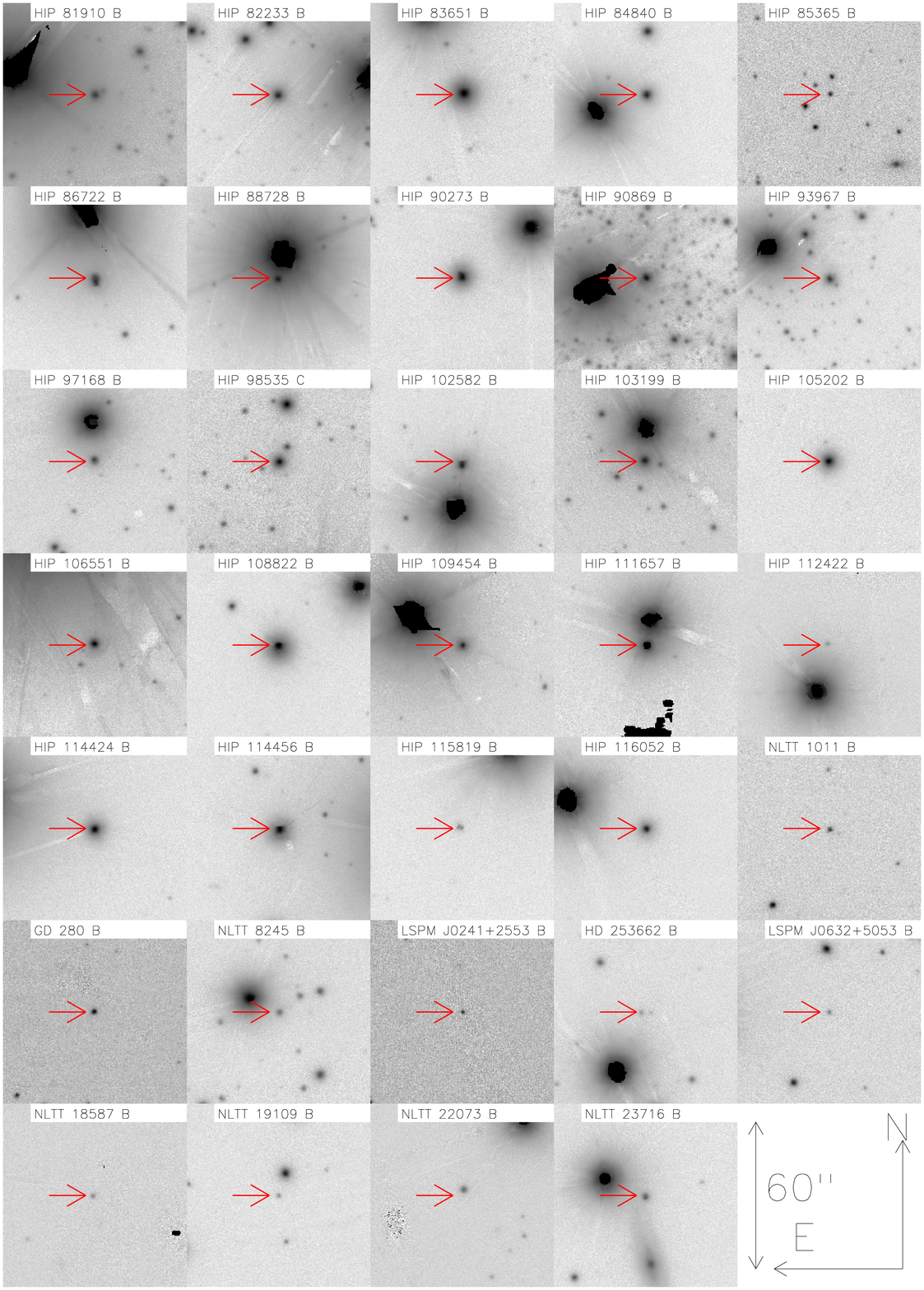}
\caption{\label{finders2} as Figure~\ref{finders1}.}
\end{center}
\end{figure}

\begin{figure}[htbp]
\begin{center}
\epsscale{0.8}
\plotone{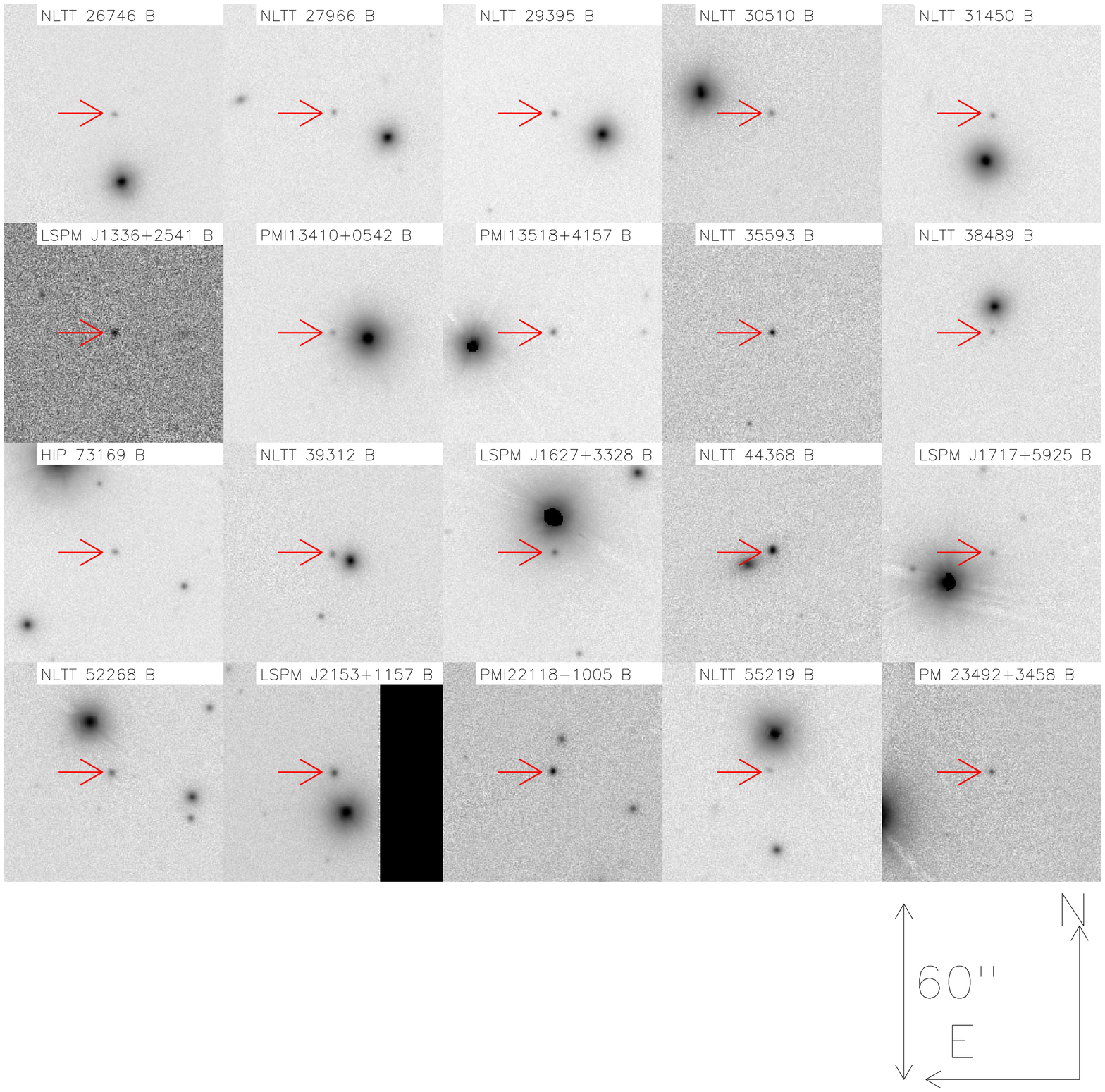}
\caption{\label{finders3} as Figure~\ref{finders1}.}
\end{center}
\end{figure}

\begin{figure}[htbp]
\begin{center}
\epsscale{1.0}
\epsscale{0.5}
\begin{tabular}{cc}
\plotone{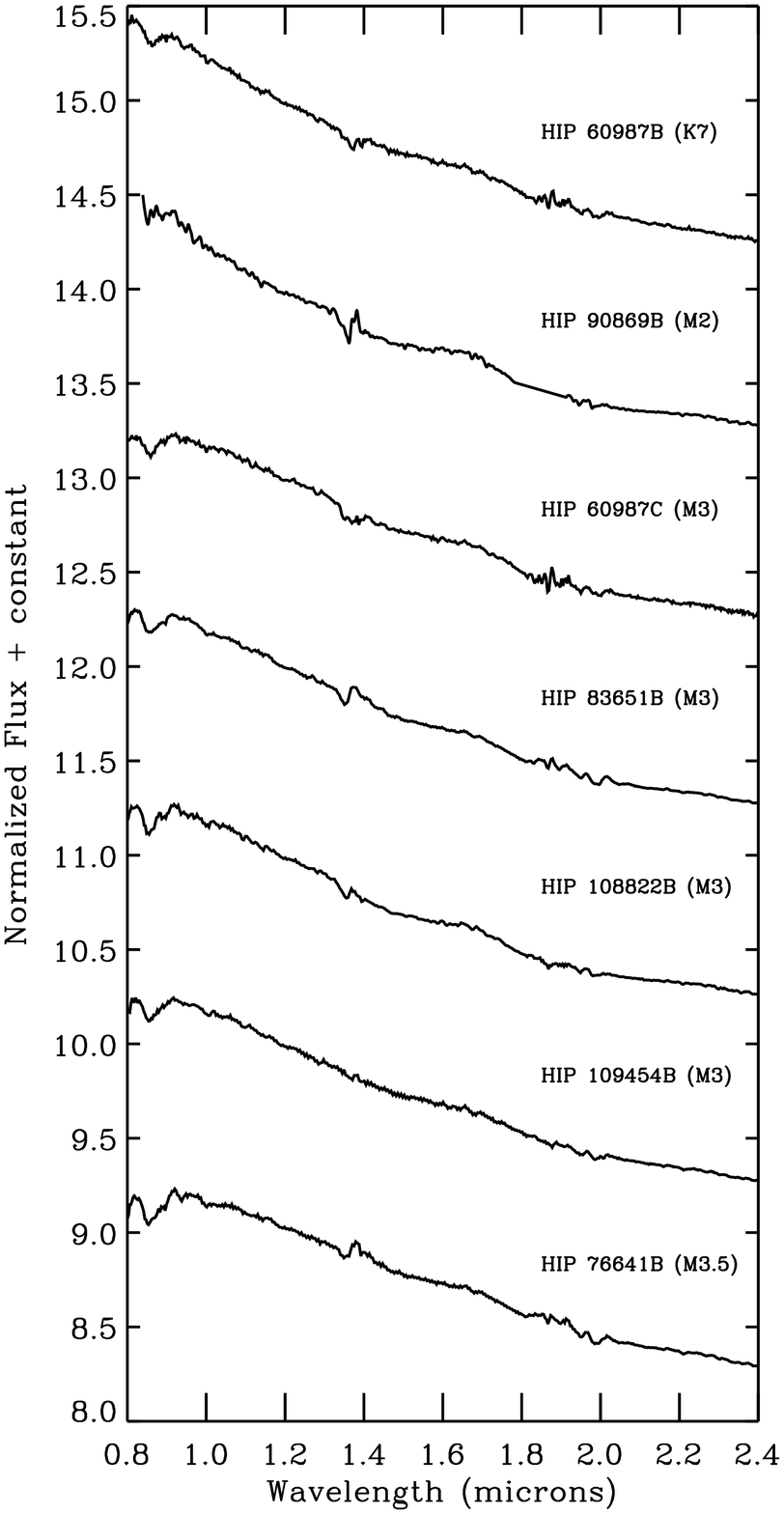}
\plotone{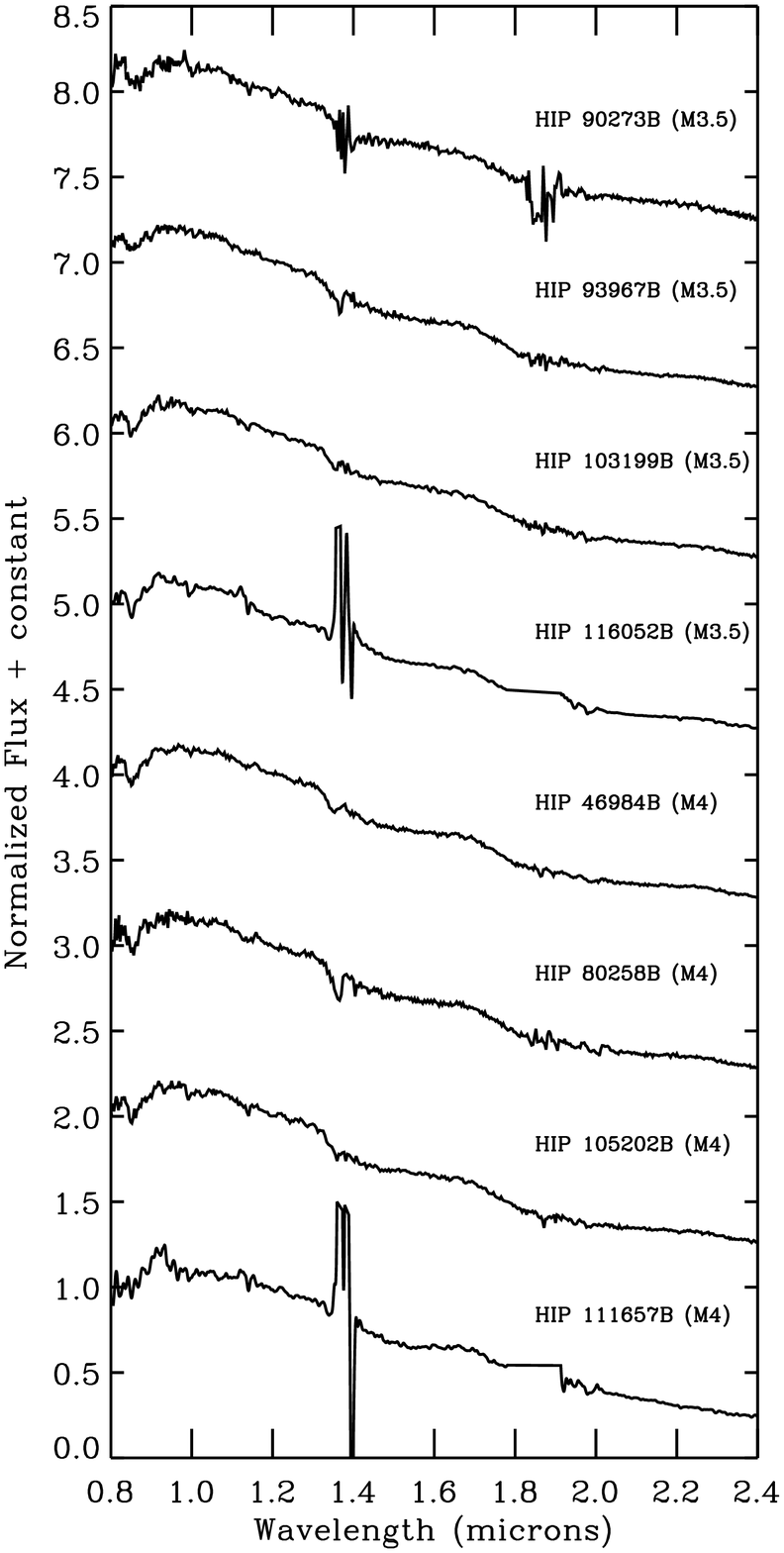}
\end{tabular}
\caption{\label{HIP_early_M_spectra} IRTF/SpeX spectra for our {\it Hipparcos} companions with spectral types from M0.5 to M4. Spectra taken with SpeX SXD mode have been gaussian smoothed to R=200. For the SXD spectra note the noisy gaps at 1.4~$\mu$m and 1.8~$\mu$m are caused by the order boundaries.}

\end{center}
\end{figure}

\begin{figure}[htbp]
\begin{center}
\epsscale{0.5}
\begin{tabular}{ccc}
\plotone{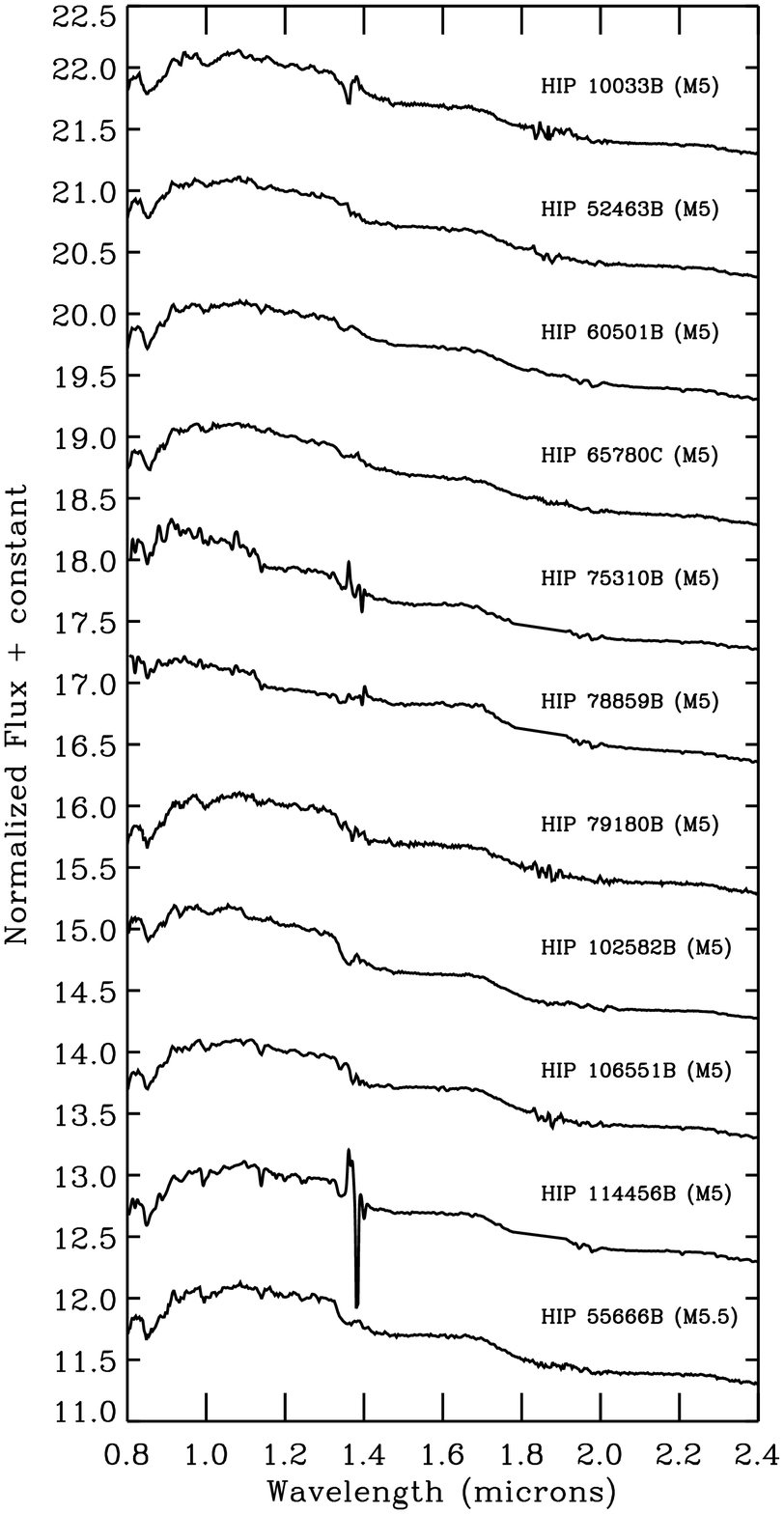}
\plotone{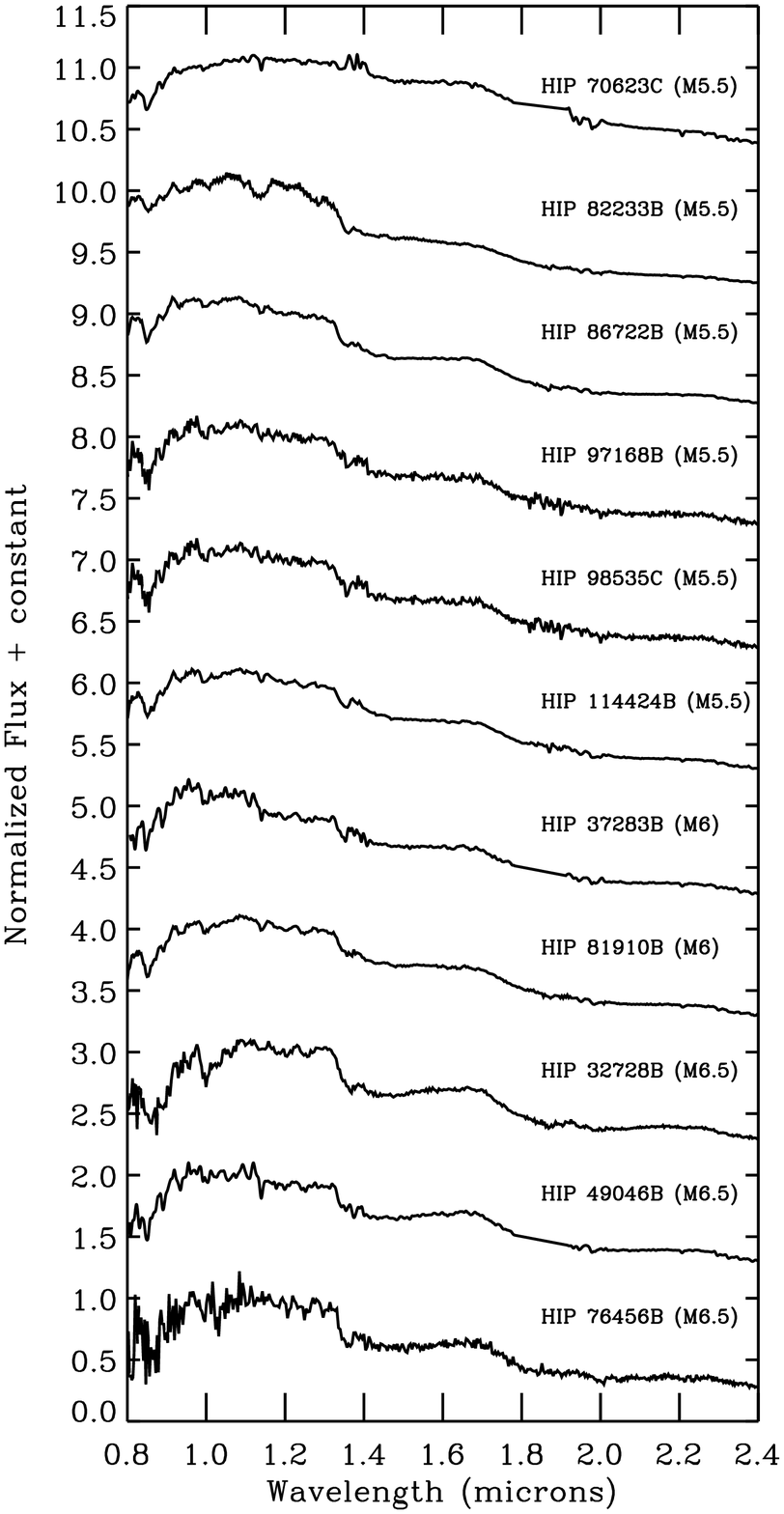}
\end{tabular}
\caption{\label{HIP_mid_M_spectra}IRTF/SpeX spectra for our {\it Hipparcos} companions with spectral types from M5 to M6.5. Spectra taken with SpeX SXD mode have been gaussian smoothed to R=200.}
\end{center}
\end{figure}

\begin{figure}[htbp]
\begin{center}
\epsscale{0.5}
\begin{tabular}{cc}
\plotone{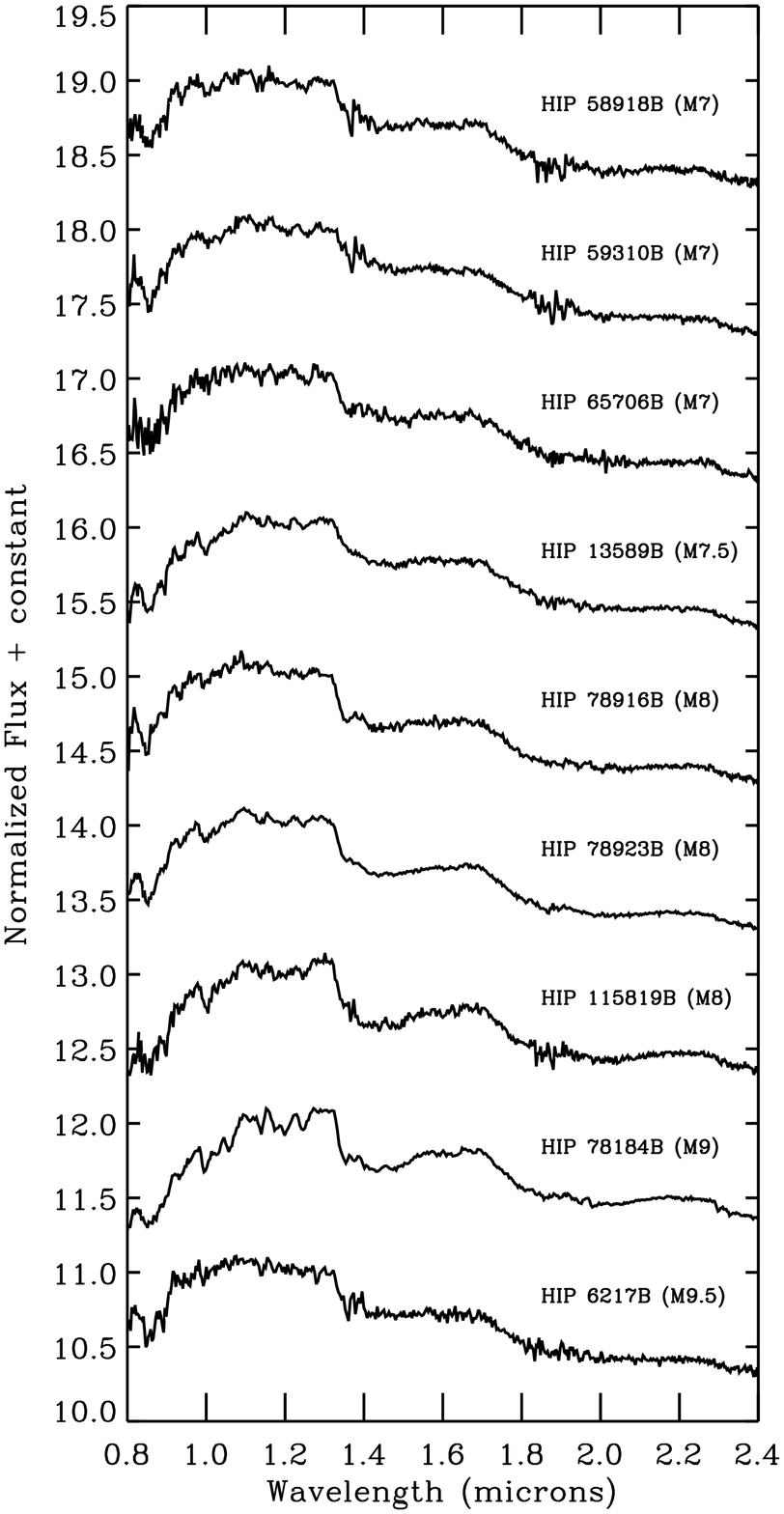}
\plotone{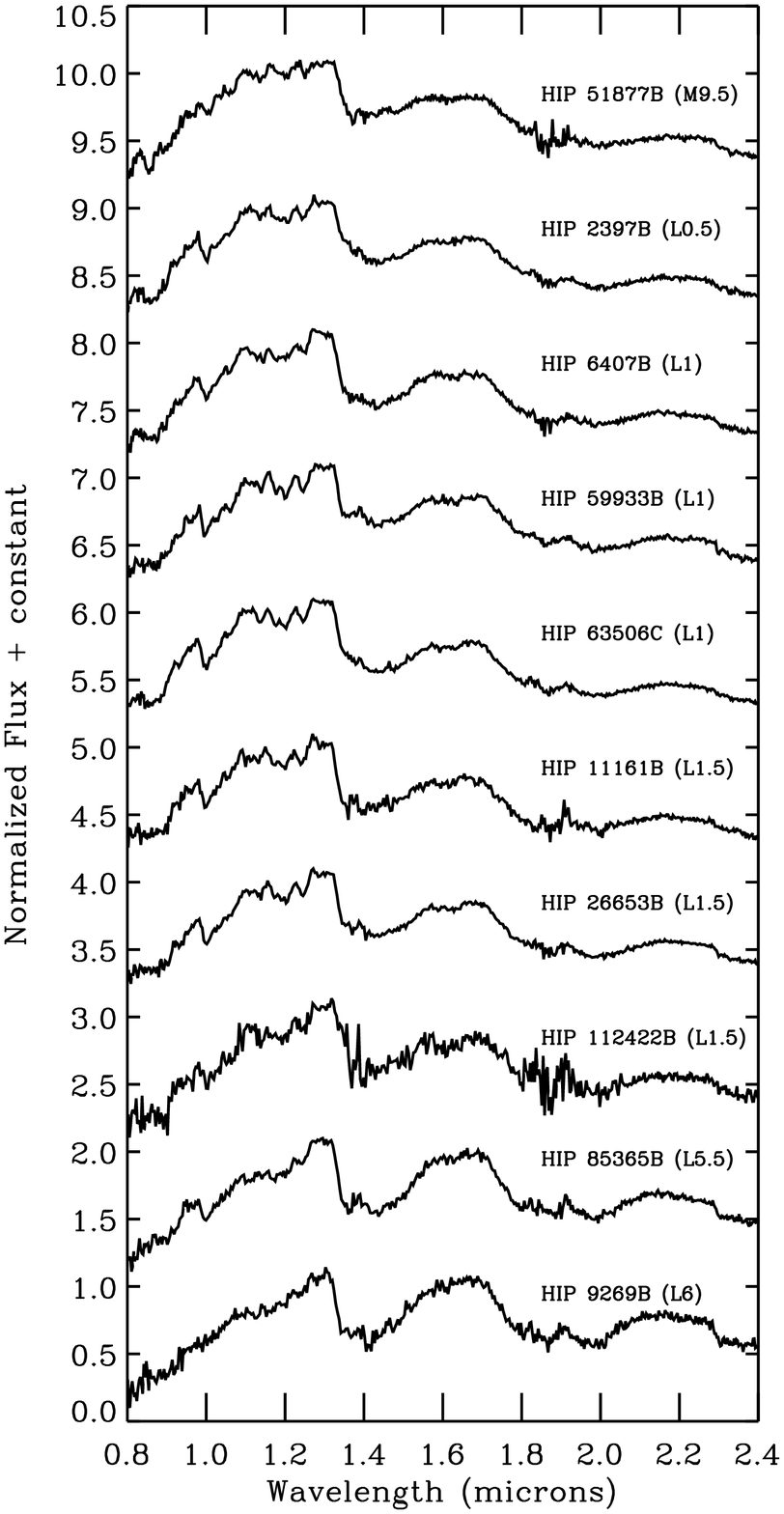}
\end{tabular}
\caption{\label{HIP_ultracool_spectra} IRTF/SpeX spectra for our ultracool {\it Hipparcos} companions with spectral types of M7 or later. Spectra taken with SpeX SXD mode have been gaussian smoothed to R=200.  Note HIP~6407~B was resolved as being an L1+T3 binary itself. See Section~\ref{HIP_6407_descript} for details.}

\end{center}
\end{figure}

In total we have spectrally typed 87 companions to nearby stars of which 56
are new discoveries. Of these, 24 of our new discoveries are L dwarf
companions, one is a late K , one is a DA white dwarf, 24 M7-M9.5 dwarfs with the remaining 37 being M0-M6.5 dwarfs. Finder charts for our new companions are shown in Figures~\ref{finders1}, \ref{finders2} and \ref{finders3}. In addition, we identified one previously known blue L~dwarf (2MASS J00150206+2959323, \citealt{Kirkpatrick2010}) as a companion to an M~dwarf (NLTT~730), and a candidate L~dwarf companion to an M dwarf (NLTT 35593) which we consider an unlikely companion to its M~dwarf primary.
\subsection{Companionship checks}
In order to assess whether our candidate companions are merely
alignments of unassociated stars, we undertook a test similar to
\cite{Lepine2007}. This consisted of taking all the objects in our input
lists ({\it Hipparcos}, LSPM/rNLTT and \citealt{Lepine2011}) and offsetting their positions by 2 degrees
in Right Ascension. We then searched for common proper motion companions to
this list of modified positions. This test should only yield non-physical (coincident) pairings. To accurately reflect the probability of our
objects being coincident contaminants, we applied the same cuts that were applied to our initial target sample. As an additional cut to exclude spuriously red objects (which we have excluded from our candidate sample by checking the objects $g_{P1}$, $r_{P1}$ and $i_{P1}$ magnitudes) we excluded objects with more than one $g_{P1}$ detection. 

In Figure~\ref{HIP_shifted}, we compare the
separation and proper motion differences for our HIP samples and the coincident
population. All objects in both our bright and ultracool companion
 samples which were followed up spectroscopically lie in portions of the plot sparsely populated by coincident
pairings. This is likely due to our approach of preferentially following-up closer companions. Figure~\ref{others_shifted} shows the results for companions to non-{\it Hipparcos} primaries and our
serendipitous discoveries. Most of our discoveries lie in a completely different area of the plot from the coincident pairings. However one object, the apparent companion to NLTT~35593, is in a region of the diagram populated by many coincident pairings, and therefore we do not consider it to be a bona-fide companion. Excluding this object, we consider it likely that less than five of our companions to {\it Hipparcos} stars will be chance alignments with background stars. For our faint non-{\it Hipparcos} primaries, at most two could be chance alignments. All of these coincident pairs would lie in the lowest proper motion bin of our sample.
\subsubsection{Photometric distance}
\label{phot_dist}
To further check the companionship of the objects, we derived photometric distances for all of our companions based on their spectral type. For objects of spectral types M6 and later, we derived the absolute magnitudes for our objects using the relations of \cite{Dupuy2012} in $J$, $H$ and $K/K_S$. For earlier-type objects, we used the SED templates of \cite{Kraus2007}, who do not quote an RMS about their relations. To estimate this we calculated the standard deviation of non-saturated 2MASS photometry of the FGK stars in \cite{Valenti2005} about the main sequence. These were found to be 0.25, 0.22 and 0.22\,mag in $J$, $H$ and $K_s$ respectively. \cite{Cruz2002} calculated fits to $J$-band absolute magnitude as a function of temperature-sensitive spectral features for M~dwarfs. They found the M~dwarf population was well-fitted by two relations, one covering early~M with an RMS of $\sim$0.2mag and the other covering late-M with an RMS of $\sim$0.35mag. This latter number is comparable with the 0.39\,mag RMS on the 2MASS $J$-band relation of \cite{Dupuy2012}. Hence we use the RMS of \cite{Dupuy2012} for objects of spectral type M5 and later and our calculated RMS for earlier-type objects. 

Next, we calculated absolutes magnitude from the \cite{Kraus2007} or \cite{Dupuy2012} relations and compared these with the object's 2MASS photometry, or if available VISTA, UKIDSS or UKIRT photometry, producing a distance estimate for each filter. Where non-2MASS near-infrared photometry was available for an object in a particular filter, we used this instead of 2MASS making use of the \cite{Dupuy2012} MKO relations for ultracool dwarfs and converting to the 2MASS system using the transformations of \cite{Carpenter2001} for comparisons to the \cite{Kraus2007} SEDs. We calculated the errors in distance caused by an uncertainty in spectral type, by the RMS of the fits and by the error on the 2MASS photometry. We then calculated the weighted mean of these distance estimates weighting only by our measurement errors i.e. the quadrature sum of the error in the photometric measurements and the propagated error in spectral typing. We then calculated a final error on this distance based on the quadrature sum of the photometric error,  intrinsic scatter and propagated error in spectral type for each band.  Hence the final quoted error includes both the effects of measurement errors and the RMS intrinsic scatter about the photometric relations. The calculated photometric distances are shown in Tables~\ref{HIPsystems} and \ref{othersystems}.  Note for our white dwarf companion (HIP~88728~B ) was saturated in Pan-STARRS\,1, is not resolved in plate images.

 We also calculated photometric distances for our primaries which lacked trigonometric parallaxes. For primaries with measured spectral types, we applied a similar process as for our secondaries. Where there was no measured spectral  distance relations of \cite{Lepine2005b} and calculated the errors in those distances in the same way as for the secondaries. For our two white dwarf primaries we used the photometric distance relations of \cite{Limoges2013}. For this we assumed a DA spectral type and a mass of 0.6\,M$_{\odot}$. \cite{Limoges2013} quote a photometric distance error of 15\,pc for their sample with approximate distances of 30\,pc. Hence we assume errors of 50\% for our white dwarf photometric distances. It appears most of our companions have photometric distances that are in good agreement with the distances to their primaries. This is shown in Figure~\ref{d_phot_plot}.

\subsection{Spectral types and ages of the primaries}
\subsubsection{Spectral types}
We searched the literature for information on our primary stars. While many had spectral types either from previous measurements, 32 were unclassified. For primaries with no published spectral type we have obtained where possible near infrared or optical spectroscopic observations(see Section~\ref{IRspec} and \ref{Optspec}). These objects are shown in Figures~\ref{HIP_primaries} and \ref{optical_primaries}. For objects with no literature spectral type or which were not observed as part of our follow-up program, we used the $V-J$ to spectral type relation of \cite{Lepine2011}. For our non-{\it Hipparcos} primaries, we use the $M_V$ magnitudes listed in \cite{Lepine2005} and \cite{Salim2003} along with $J$ magnitudes from 2MASS. For a single object which was too blue for this relation to  be valid (HIP~111657), we used the primary's 2MASS photometry and the spectral energy distributions (SED) of \cite{Kraus2007} to estimate spectral type. Table~\ref{HIPprimary_kin} and Table~\ref{othersprimary} show details of our primary stars. 

 For objects observed with SNIFS, spectral types for NLTT 730, LSPM J2153+1157, and NLTT 22073 were determined following the methods outlined in \citet{Lepine2013}. Specifically, we measured the strengths of the TiO and CaH features and then compared to stars from \citet{Reid1995}. We also matched each spectrum by eye to templates from \citet{Bochanski2007} using the IDL spectral typing suite of \cite{Covey2007}. Metallicities were determined following the methods of \citet{Mann2013}, which provide empirical relations between visible-wavelength features and metallicities for late-K to mid-M dwarfs. Errors in metallicities were calculated considering both errors in the \citet{Mann2013} calibration and measurement errors. We classified NLTT~1011 using our Mayall Telescope data and the \citet{Covey2007} HAMMER indices, resulting in a type of K5. Our other Kitt Peak targets, NLTT~38489  and HIP~84840~B were visually compared to the spectral templates of \citet{Bochanski2007} with best matches of M3  and M4 respectively, these subtype was confirmed by index measurements using the method of \cite{Lepine2013}  for NLTT~38489 while HIP~84840~B was classified as an M3.5. We adopt our visual comparisons along with a spectral typing error of half a subclass for these two objects. None of our observed objects showed emission in H$\alpha$. One of our SNIFS targets GD~280 was listed as a candidate white dwarf by \cite{Giclas1967}. Our spectrum shows clear Balmer line absorption; hence we classify it as a DA white dwarf. We do not have a spectrum for LSPM~J0241+2553. The resulting spectral types and metallicites for the objects observed with SNIFS are listed in Table~\ref{othersprimary} and the spectra are shown in Figure~\ref{optical_primaries}.

\subsubsection{Age Estimates}

 Twenty-three of our primary stars have age estimates listed in the literature. For remaining objects with no published ages, we used archival Ca H and K emission and where this was found applied the age-activity relation of \cite{Mamajek2008}. The majority of our primaries had no such data, only one object which did had no age derived from such emission in the literature. For the remaining 54 objects with no Ca H and K measurements, we set an approximate upper age limit of $\sim$10\,Gyr based on their disk-like kinematics. 

To set a lower limit for the handful of M-dwarf primaries where we had an optical spectrum, we used the object's lack of H$\alpha$ and the activity lifetime of \cite{West2008}. Here we used the 1$\sigma$ lower limit for the activity age taking into account that our spectral types have an uncertainty of 0.5 spectral types. For all our objects without H$\alpha$ or Ca H and K data, we searched the ROSAT Faint Source catalog \citep{Voges2000} with a matching radius of 30\arcsec. We identified four objects as having weak X-ray emission; the rest we assumed the flux was below the limiting flux quoted by \cite{Schmitt1995}. We then used the distances to these objects along with the counts to flux conversion of \cite{Schmitt1995} to estimate the X-ray luminousity. We converted this to the X-ray to bolometric luminosity ratio ($R_x$) using an $L_{bol}$ calculated from Tycho photometry \citep{Hog2000}, the colour relations of \cite{Mamajek2002}, and the bolometric corrections of \cite{Pecaut2013}. This was then applied to the age-to-X-ray activity relation of \cite{Mamajek2008} to obtain lower age limits. Note this X-ray relation has not been calibrated beyond mid-K spectral types, hence we used objects' lack of X-ray emission to set a lower age limit of 300\,Myr for objects of this spectral type. This is the age below which low mass stars show significant X-ray emission \citep{Shkolnik2009}.
\section{Discussion}

\subsection{Comparison with the field ultracool wide binary population}
Our sample contains 24 new L-dwarf companions to main sequence stars
(including the previously typed, but unrecognized L7 companion to NLTT 730 but not
the unlikely L2 companion to NLTT~35593). Additionally we identified
 21 new wide M-dwarf companions with spectral types of M7 or later 
and typed two previously proposed companions.

In order to study the wide, ultracool (spectral type $\geq$M7) binary population, we compiled a list of
wide ultracool companions to stars. We began with the compilation of
\cite{Faherty2010} (excluding the companion to NLTT~20346 which \citealt{Dupuy2012} concluded was not physically related due to its high proper motion difference from its primary) and added spectroscopically confirmed objects from the
literature discovered since then. We do not include objects that have been identified as candidate binaries but which lack spectral types such as from the studies by \cite{Deacon2009a}, \cite{Dhital2010} and \cite{Smith2013}. Figure~\ref{all_binaries} shows the spectral
type of the secondary plotted against projected separation ($r_{AU}$). At first glance, there is an apparent scarcity of T dwarf companions wider than 3000~AU. In fact, this is due to previous efforts focussing on identifying close companions and from the known population being drawn from a series of heterogeneous surveys. Figure~\ref{binaries_hist} shows a histogram comparing
the combined contribution of this work, \cite{Deacon2012} and
\cite{Deacon2012a} to the total number of companions. This paper's
contribution is most significant beyond $\log_{10}r_{AU}=$3.5 where we have
doubled the population of L dwarf companions. In total we have increased the wide ($>$300~AU) L dwarf companion population by 82\% and doubled the number of ultracool M dwarf companions in the same range. While the L~dwarf population
exhibits an approximately flat distribution in $\log_{10}r_{AU}$, the T~dwarf
companion population peaks between $\log_{10}r_{AU}=$3.0 and 3.5. However any
claim of a preferred separation for T~dwarf companions (or of a log-flat distribution for L dwarfs) should be treated with caution as the sample is drawn from a disparate
set of surveys. Similarly the apparent cut-off above 10,000~AU is likely
due to incompleteness in surveys (including our own) and the difficulty of
disentangling the widest binaries from coincident alignments from field
stars. In a future paper we aim to take the final results of our survey, model our selection biases and determine the true separation distribution of the ultracool companion population.

In order to determine the
stability of the population of systems containing a wide, ultracool companion, we estimated the total mass of each system. Where an object had a mass quoted in the
literature, we used that value. For known L and T dwarfs with no quoted mass
estimates and our new L and T companions, we used a value of 0.075~M$_{\odot}$, in essence making our total
mass estimates for these systems upper limits. For A-M stars with no mass in
the literature, we converted their spectral type to mass using the relations
of \cite{Kraus2007}\footnote For the LSPM J1202+0742 ABC system \citep{Smith2013}, we estimated the spectral type of the brighter component based on their $V-J$ colour (M0, LSPM~J1202+0742N) or 2MASS photometry (M1, LSPM~J1202+0742S). We then used these to determine the total mass of the system. The results are plotted in Figure~\ref{binaries_mtot}
along with the maximum separation vs.\ total mass relations suggested by
\cite{Close2003} and \cite{Reid2001a}. It is clear that a substantial number
of late-type companions lie outwith both of these suggested maximum
boundaries and are hence loosely bound. \cite{Dhital2010} used a simple model
of interactions in the Galactic disk (based on \citealt{Weinberg1987}) to
calculate the typical maximum separation for a given system age and total
mass. This is also plotted in Figure~\ref{binaries_mtot} for a number of
different ages. It is clear that, though loosely bound, very few of the widest
systems would be disrupted over the lifetime of the Galactic disk.

\subsection{Interesting individual systems}
\subsubsection{HIP 6407Bab}
\label{HIP_6407_descript}
We initially classified HIP~6407B as an L0 based on visual comparison to standards. However the object has significantly stronger water absorption features suggestive of an object two or three subtypes later. As this may result from the contribution of an unresolved additional component, we observed this object on 2013~October~14~UT at the Keck~II telescope using
the facility near-infrared camera NIRC2 with laser guide star adaptive
optics \citep[LGS AO;][]{Wizinowich2006,VanDam2006}.
We kept the LGS centered in NIRC2's narrow field-of-view camera while
we obtained dithered images of the target in the $Y_{\rm NIRC2}$,
$J_{\rm MKO}$, $H_{\rm MKO}$, and $CH_4s$ bandpassess. The wavefront
sensor recorded flux from the LGS equivalent to a $V \approx
9.6$--9.8~\,mag star. The primary star HIP~6407 was used as the
tip-tilt reference star, and the lower bandwidth sensor monitoring
this source recorded flux equivalent to a $R \approx 7.6$\,mag star.
Our procedure for reducing and analyzing Keck LGS data is described in
detail in our previous work \citep[e.g.,][]{Liu2006,Dupuy2009a}. To summarize briefly, we measure binary
parameters by fitting three-component Gaussians to determine the
position and flux of each binary component, and we derive
uncertainties by computing the scatter among individual dithered
images. We used the NIRC2 astrometric calibration from
\cite{Yelda2010}, which includes a correction for the
nonlinear distortion of the camera and has a pixel scale of
$9.952\pm0.002$\,mas~per~pixel and an orientation for the detector's
$+y$-axis of $+0\fdg252\pm0\fdg009$ east of north. We resolved HIP 6407B as a 0.131,\arcsec (7.4\,AU) binary. Table~\ref{LGStable} gives the
binary parameters we measured in each bandpass along with the weighted
average of the separation ($\rho$) and position angle (P.A.) values. Given the objects' separation and maximum total mass of 0.16\,M$_{\odot}$ (as it consists of two substellar objects), the likely period of this system is $>$300\,years, making it a poor mass-age benchmark.

 We performed spectral decomposition analysis on HIP~6407B using the
method described in Section~5.2 of \citet{Dupuy2012}.
Briefly, we started with all possible pairs of the 178 IRTF/SpeX prism
spectra from the library of \citet{Burgasser2010}. For each
template pairing, we determined the scale factors needed to minimize
the rms deviation from our observed spectrum. We then computed the
$\chi^2$ of our Keck LGS AO flux ratios in $J$, $H$, and $CH_4s$ bands
compared to the flux ratios computed for each pairing. We excluded
pairings that significantly disagreed with our measured flux ratios,
$p(\chi^2) < 0.05$, and examined the remaining best pairings to
determine the component spectral types. Our final spectral types of
L$1.0\pm0.5$ and T$3\pm1$ account for the full range of spectral
templates that gave equally good fits to our combined light spectrum.
We estimated the flux ratio in $K$-band as well as flux ratios in the
2MASS photometric system by taking the mean and rms of each flux ratio
among the ensemble of best-fit template pairings. Figure~\ref{HIP_6407B} shows the
best match to our spectrum, which is provided by the templates.

The best match to our spectrum is provided by the templates
DENIS-P~J170548.3$-$051645 \citep[L1;][]{Burgasser2010,Allers2013} and SDSS~J120602.51+281328.7
\citep[T3;][]{Chiu2006,Burgasser2010} scaled to each other by the
magnitudes shown in the lower panel of Figure~\ref{HIP_6407B}, giving magnitude differences of $d_J=2.26\pm0.05$ mag and $d_H=2.51\pm0.05$ mag. We adopt spectral types of L$1\pm1$
and T$3\pm1$ for the components of HIP~6407B based on the typical
uncertainty in infrared types for L dwarfs and the range of good
matching templates for the secondary. Note that the template library
from \citet{Burgasser2010} nominally only includes objects later
than L0, which barely encompasses the spectral type of the proposed primary component.
However, Figure~\ref{HIP_6407B} shows that the L1 template gives a very good match
to the data, so the primary is not likely to be of much earlier type.

\subsubsection{HIP 70623 (HD 126614)}
HIP 70623 (HD~126614) is a K0 star with a $M \sin i = 0.38\,M_{Jup}$ giant planet in a 3.41-yr orbit
\citep{Howard2010}. There is also an additional M dwarf component in the
system with a separation of 0.5\arcsec\, \citep{Howard2010} detected by adaptive optics imaging. The wide
M5.5 companion we recover was identified by \cite{Gould2004a} as a companion but has no previously published spectral type. Despite this object being identified as a companion before the \cite{Howard2010} identification of the closer companion, it has the designation HIP 70623/HD 126614 C.
\subsubsection{HIP 115819 (VZ Piscium)}	
VZ Piscium is a marginal-contact eclipsing binary with a period of 0.261~days
\citep{Hrivnak1995} classified as a W-UMa type variable. As angular momentum
transfer to additional components can tighten an inner binary and lead to
contact, \cite{Rucinski2007} observed VZ Piscium as part of an adaptive optics
search for companions to contact binaries. We have identified an M8 companion at a separation of 30\arcsec. This 
fell outside the 36\arcsec$\times$36\arcsec\ field of view of \cite{Rucinski2007}. Note that
\cite{Qian2004} suggest a closer companion to the binary may be causing a
variation in the lightcurve with a period of $\sim$25~years. However, no close
companion was identified by \cite{Rucinski2007}, and our companion is too wide to introduce such a short period variation.
\subsubsection{HIP 103199 (HD 335289)}
HIP 103199 (spectral type G5) is listed as runaway star by \cite{Tetzlaff2011}. This is based on an estimated age of $\sim$46$\pm$23~Myr and the star's space motion. \cite{Fujii2011} hypothesize that such objects are the result of a 3-body interaction that results in the formation of a runaway star and a close binary. HIP 103199 has a wide ($\geq$646~AU) binary companion found by \citealt{Lepine2007}, which we classify as M3.5. We converted our UKIRT photometry to the CIT system using the conversions of \cite{Carpenter2001} in order to calculate the properties of the companion using evolutionary models. Using the {\it Hipparcos} distance to the primary of 59.5$^{+5.1}_{-4.3}$~pc, we derive an absolute K-band magnitude of 7.42$\pm0.18$\,mag for the companion. Using the evolutionary models of \cite{Baraffe1998} and the spectral type--effective temperature scale of \cite{Kenyon1995}, we derive K-band absolute magnitudes of 6.18$\pm$0.25\,mag at 50~Myr and 7.16$\pm$0.17\,mag at 300\,Myr for an M3.5 where the quoted error is calculated from our half a subtype classification uncertainty. Based on this we suggest that the companion to HIP 103199 (HD 335289) is not over-luminous due to youth and it is likely that this system is not as young as suggested by \cite{Tetzlaff2011}.

\subsubsection{HIP~60987~B/C}
During the spectroscopic observations of our companion to HIP~60987 (spectral type M3, separation 19\arcsec), we identified that the primary itself was a visual double. A spectrum of this other companion (spectral type K7, separation $\sim 5$\arcsec) was also obtained. As both companions had previous listings in the Washington Double Star catalog, we used the previously existing designations for the companions.

\subsubsection{NLTT~38489~A/B}
 The companion to NLTT~38489 stands out as having a photometric distance that
 is in particularly poor agreement with its primary. However this object is
 only 6.7\,\arcsec\, away from its primary. Hence the photometry used to
 estimate the companions photomtric distance may be unreliable.
\subsubsection{LSPM J0241+2553~A/B}
 We do not have a spectrum for LSPM J0241+2553~A. To characterise this object we examined its reduced proper motion diagram placement compared to Figure~4 of \cite{Limoges2013} and concluded that this object was likely a white dwarf.
\subsubsection{HD 253662~A/B}
 HD~253662~A is listed in Simbad as a G8 subgiant. We estimated this object's photometric distance using the same mechanism as for other photometric distances using the G8 spectral type, the object's 2MASS magnitudes and the absolute magnitudes quoted in \cite{Kraus2007}. As this object is a subgiant and hence more luminous than the dwarfs used to calibrate the \cite{Kraus2007} absolute magnitudes, we quote our 1$\sigma$ lower error bound as a minimum distance.
\subsubsection{The possible companion to NLTT~35593}
The candidate companion to NLTT~35593 has a very similar photometric distance (65.8$\pm_{10.7}^{12.8}$\,pc) to its primary (63.0$\pm_{14.6}^{19.1}$\,pc). However its wide separation (1106\arcsec) puts it in a similar region of Figure~\ref{others_shifted} populated by chance alignments. Hence we consider it unlikely to be a bona-fide companion. Radial velocity or parallax measurements will be required to assess whether it is a true binary.

\section{Conclusions}
We have presented 61 newly discovered companions  to known nearby stars, with 25 of these being previously unknown L dwarf companions. With the addition of the spectral classification of previously known objects, we have characterised  a total of 88 wide, common proper motion companions to nearby stars. We have  increased the sample of late~M companions with projected separations greater than $\sim$300~AU  by 96\% and increased the number of L~dwarf companions in the same separation range by 82\% . Examination of our discoveries and the previously known wide ultracool companion population indicates that although many of the systems are loosely bound, they are unlikely to be disrupted over several Gyr. This paper provides a large sample of wide, ultracool companions to stars, which are excellent laboratories for testing models of substellar evolution and atmospheres. Additionally our late-type companions provide an opportunity to extend metallicity determinations for M~dwarfs to cooler temperatures \citep{Mann2014}.

\acknowledgments
The Pan-STARRS1 Surveys (PS1) have been made possible through contributions of the Institute for Astronomy, the University of Hawaii, the Pan-STARRS Project Office, the Max-Planck Society and its participating institutes, the Max Planck Institute for Astronomy, Heidelberg and the Max Planck Institute for Extraterrestrial Physics, Garching, The Johns Hopkins University, Durham University, the University of Edinburgh, Queen's University Belfast, the Harvard-Smithsonian Center for Astrophysics, the Las Cumbres Observatory Global Telescope Network Incorporated, the National Central University of Taiwan, the Space Telescope Science Institute, the National Aeronautics and Space Administration under Grant No. NNX08AR22G issued through the Planetary Science Division of the NASA Science Mission Directorate, the National Science Foundation under Grant No. AST-1238877, the University of Maryland, and Eotvos Lorand University (ELTE). 
The authors would also like to thank Bill Golisch, Dave Griep, and Eric Volqardsen for assisting with the IRTF observations. This research has benefited from the SpeX Prism Spectral Libraries, maintained by Adam Burgasser at http //www.browndwarfs.org/spexprism. This publication makes use of data products from the Two Micron All Sky Survey, which is a joint project of the University of Massachusetts and the Infrared Processing and Analysis Center/California Institute of Technology, funded by the National Aeronautics and Space Administration and the National Science Foundation. This research has benefited from the M, L, and T dwarf compendium housed at DwarfArchives.org and maintained by Chris Gelino, Davy Kirkpatrick, and Adam Burgasser. M.C.L. and E.A.M. and were supported by NSF grants AST09-09222 (awarded to M.C.L.) and AST-0709460 (awarded to E.A.M.). E.A.M. was
also supported by AFRL Cooperative Agreement FA9451-06-2-0338. This publication makes use of data products from the Wide-field Infrared Survey Explorer, which is a joint project of the University of California, Los Angeles, and the Jet Propulsion Laboratory/California Institute of Technology, funded by the National Aeronautics and Space Administration.The United Kingdom Infrared Telescope is operated by the Joint Astronomy Centre on behalf of the Science and Technology Facilities Council of the U.K. 
This paper makes use of observations processed by the Cambridge Astronomy
Survey Unit (CASU) at the Institute of Astronomy, University of Cambridge.
 The authors would like to thank Mike Irwin and the team at CASU for making the reduced WFCAM data available promptly and Tim Carroll, Thor Wold, Jack Ehle and Watson Varricatt for assisting with UKIRT observations. This research has made use of the SIMBAD database,
operated at CDS, Strasbourg, France. The VISTA Data Flow System pipeline processing and science archive are described in \cite{Irwin2004} and \cite{Hambly2008}. We have used data from the 1st data release. This paper makes use of the Topcat software package \citep{Taylor2005}. This research has made use of the Washington Double Star Catalog maintained at the U.S. Naval Observatory. We thank Luca Casagrande, Jackie Faherty, Adam Kraus, Eddie Schlafly, and Josh Schlieder for helpful discussions and our referee S\'{e}bastien L\'{e}pine for many helpful comments which improved the manuscript. Finally, the authors wish to recognize and acknowledge
the very significant cultural role and reverence that the summit of Mauna Kea has always had
within the indigenous Hawaiian community. We are most fortunate to have the opportunity to
conduct observations from this mountain.\\
{\it Facilities:} \facility{IRTF (SpeX)}, \facility{PS1}, \facility{UKIRT (WFCAM)}, \facility{UH:2.2m (SNIFS)}

\bibliography{ndeacon}
\bibliographystyle{apj}

\begin{figure}[htbp]
\begin{center}
\epsscale{0.5}
\plotone{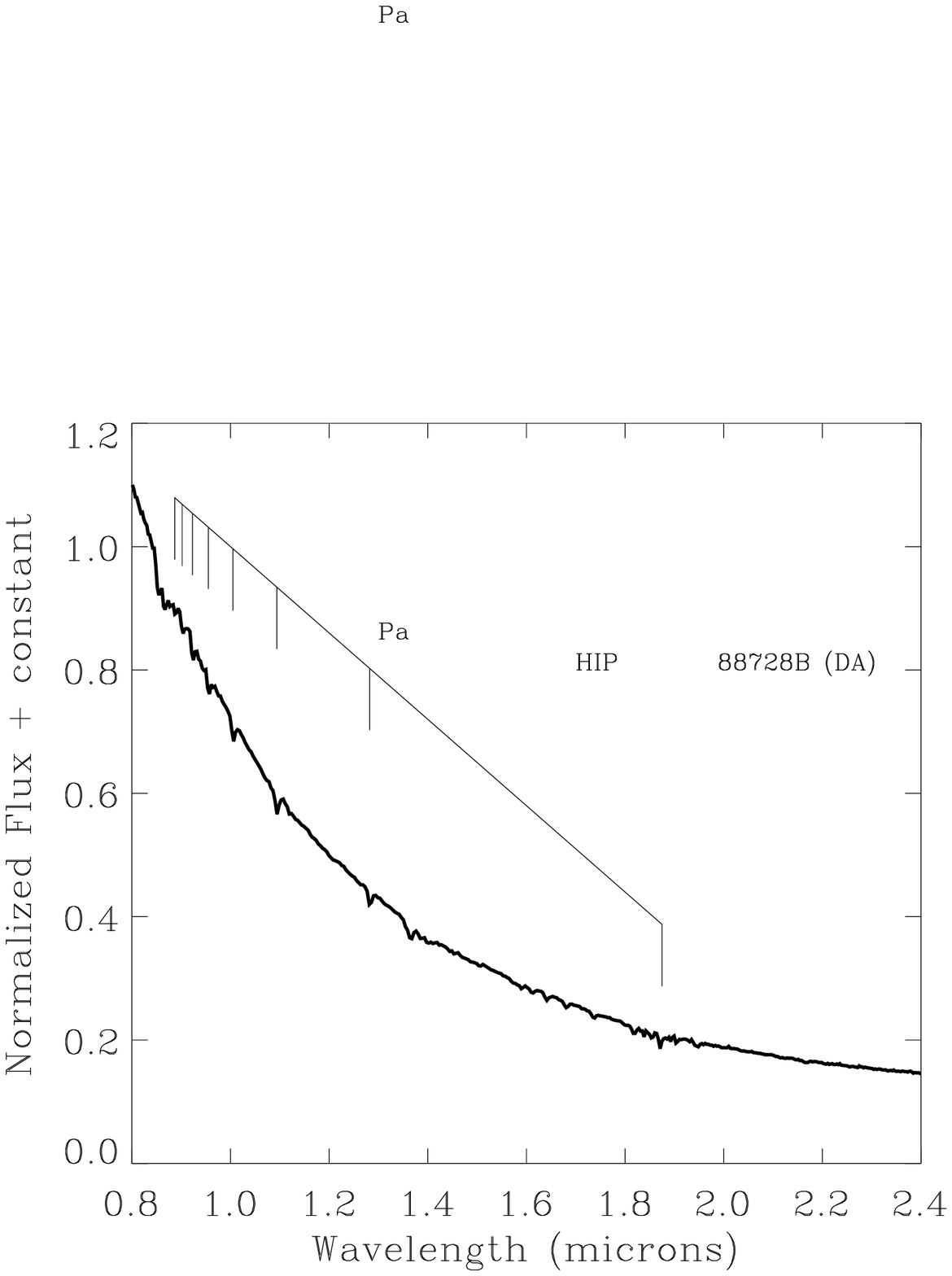}
\caption{\label{HIP_wd_spectra} IRTF/SpeX spectra for our white dwarf {\it Hipparcos} companions. The wavelengths of the Paschen lines are also shown.}

\end{center}
\end{figure}

\begin{figure}[htbp]
\begin{center}
\epsscale{0.4}
\begin{tabular}{cc}
\plotone{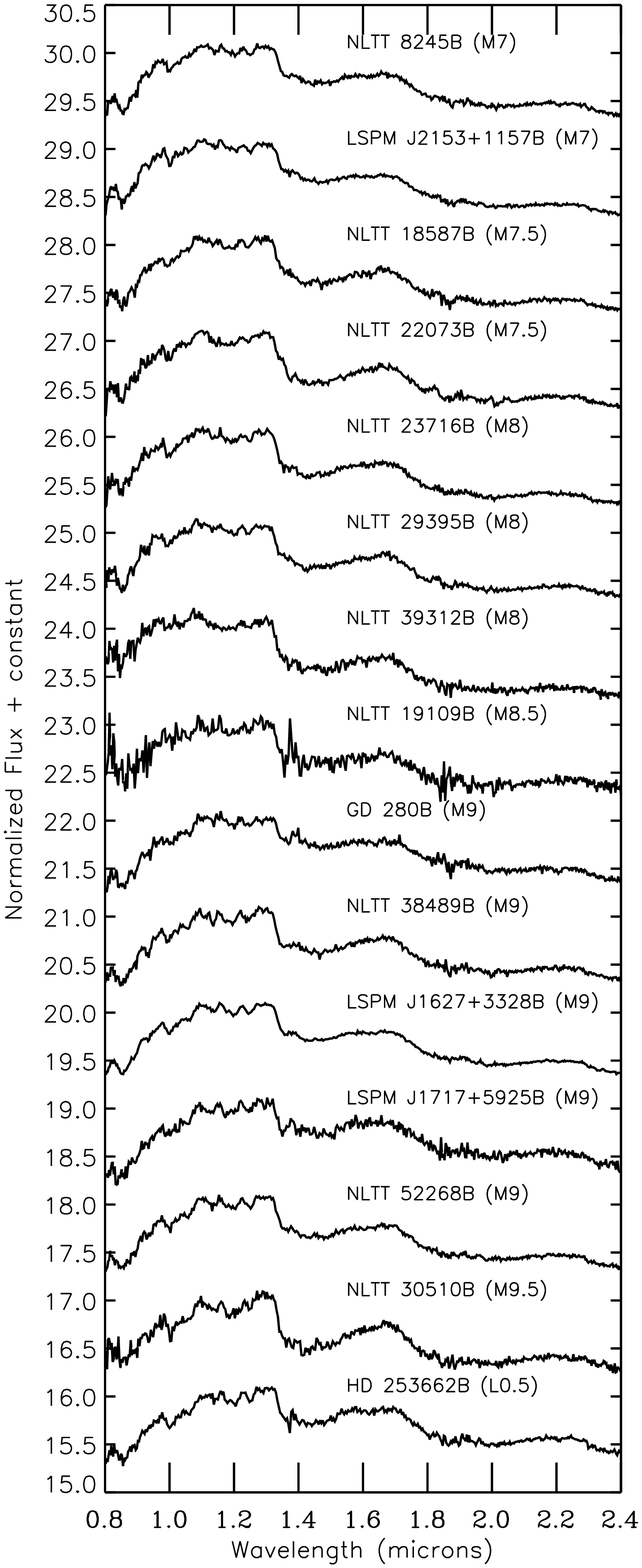}
\plotone{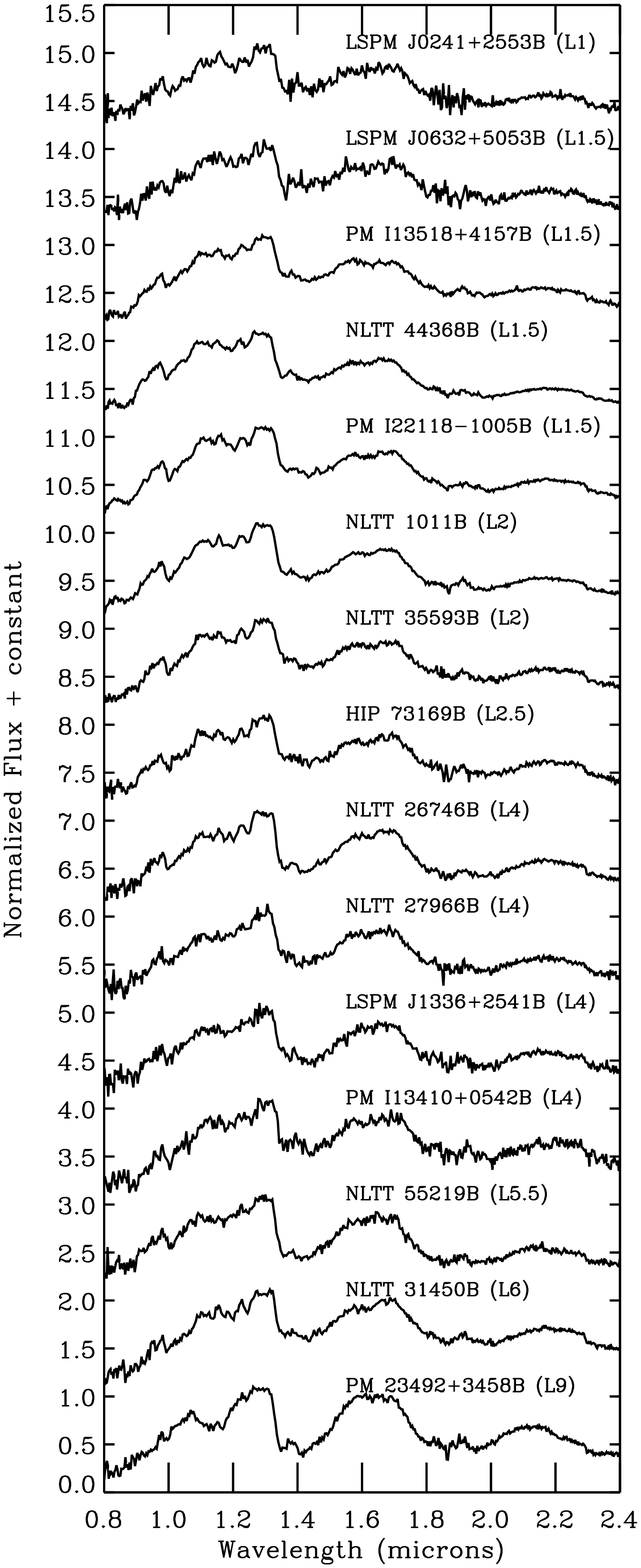}
\end{tabular}
\vspace{-10mm}
\caption{\label{other_spectra}IRTF/SpeX spectra for our ultracool companions
 discovered serendipitously or by searching for wide companions to faint non-{\it Hipparcos} primaries.}
\end{center}
\end{figure}
\begin{figure}[htbp]
\begin{center}
\epsscale{0.5}
\begin{tabular}{cc}
\plotone{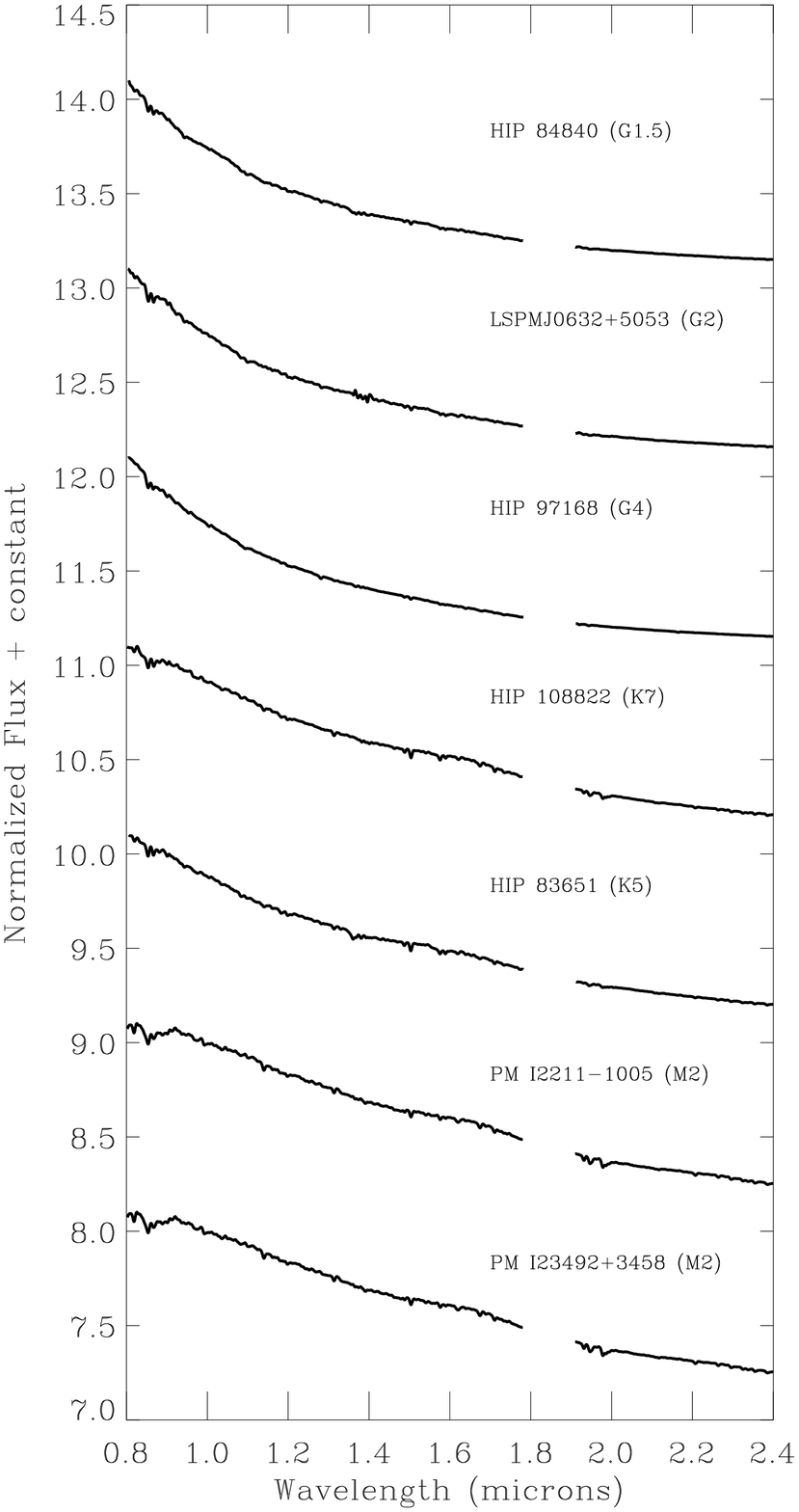}
\plotone{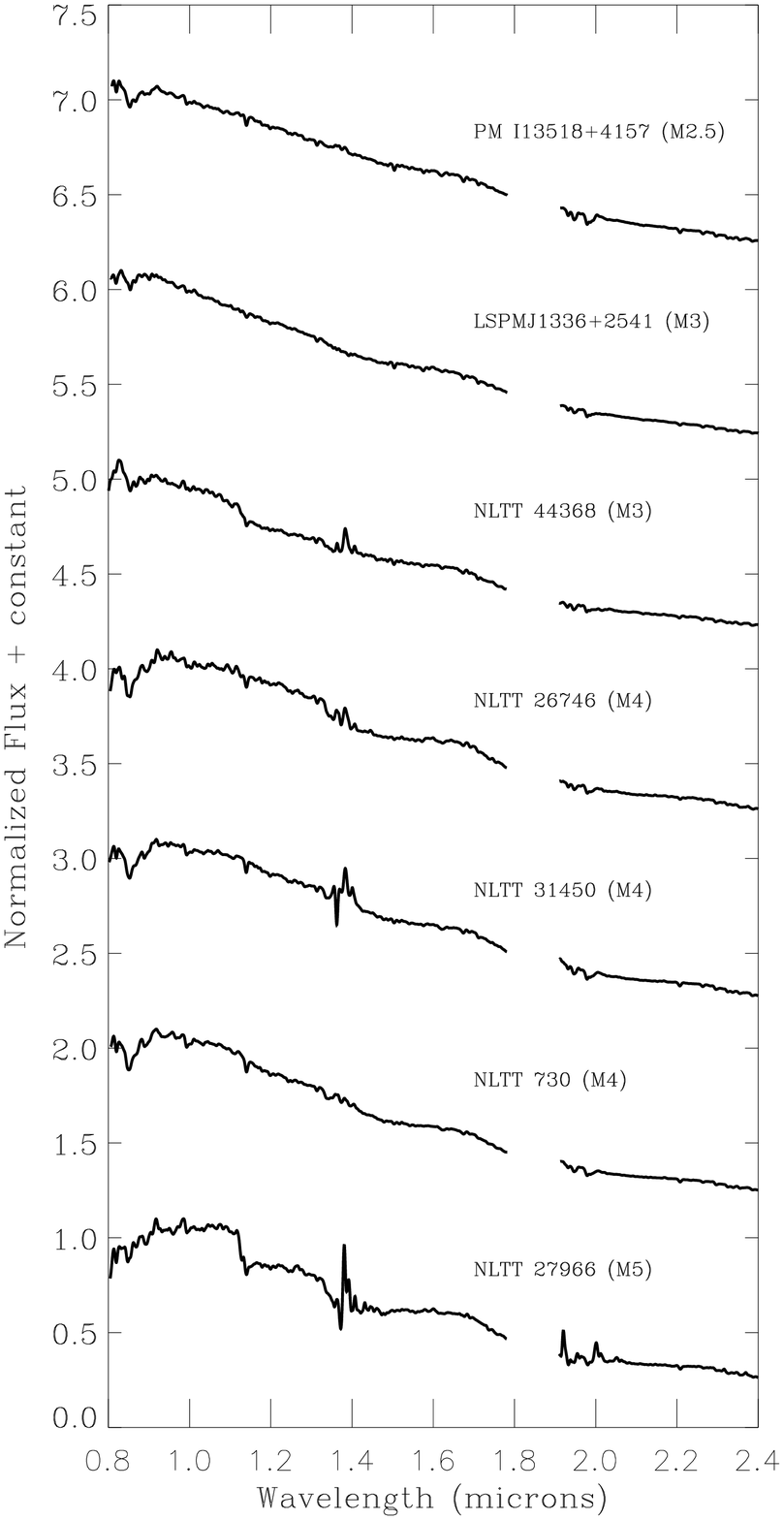}
\end{tabular}
\caption{\label{HIP_primaries} IRTF/SpeX spectra for the primaries of our companions that had no spectral type in the literature. These spectra were taken with SpeX SXD mode and have been gaussian smoothed to R=200.}

\end{center}
\end{figure}

\begin{figure}[htbp]
\begin{center}
\epsscale{0.5}
\begin{tabular}{cc}
\plotone{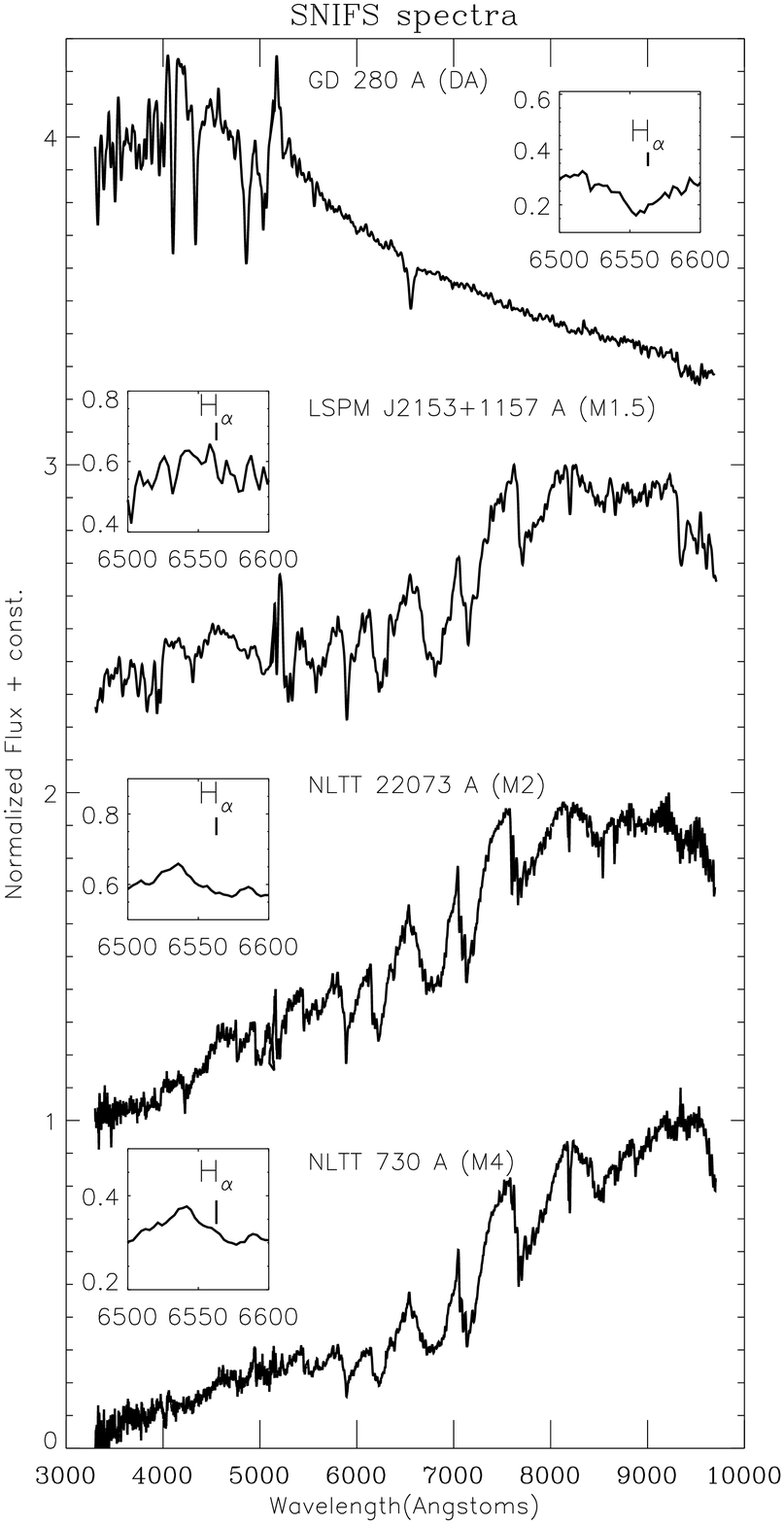}
\plotone{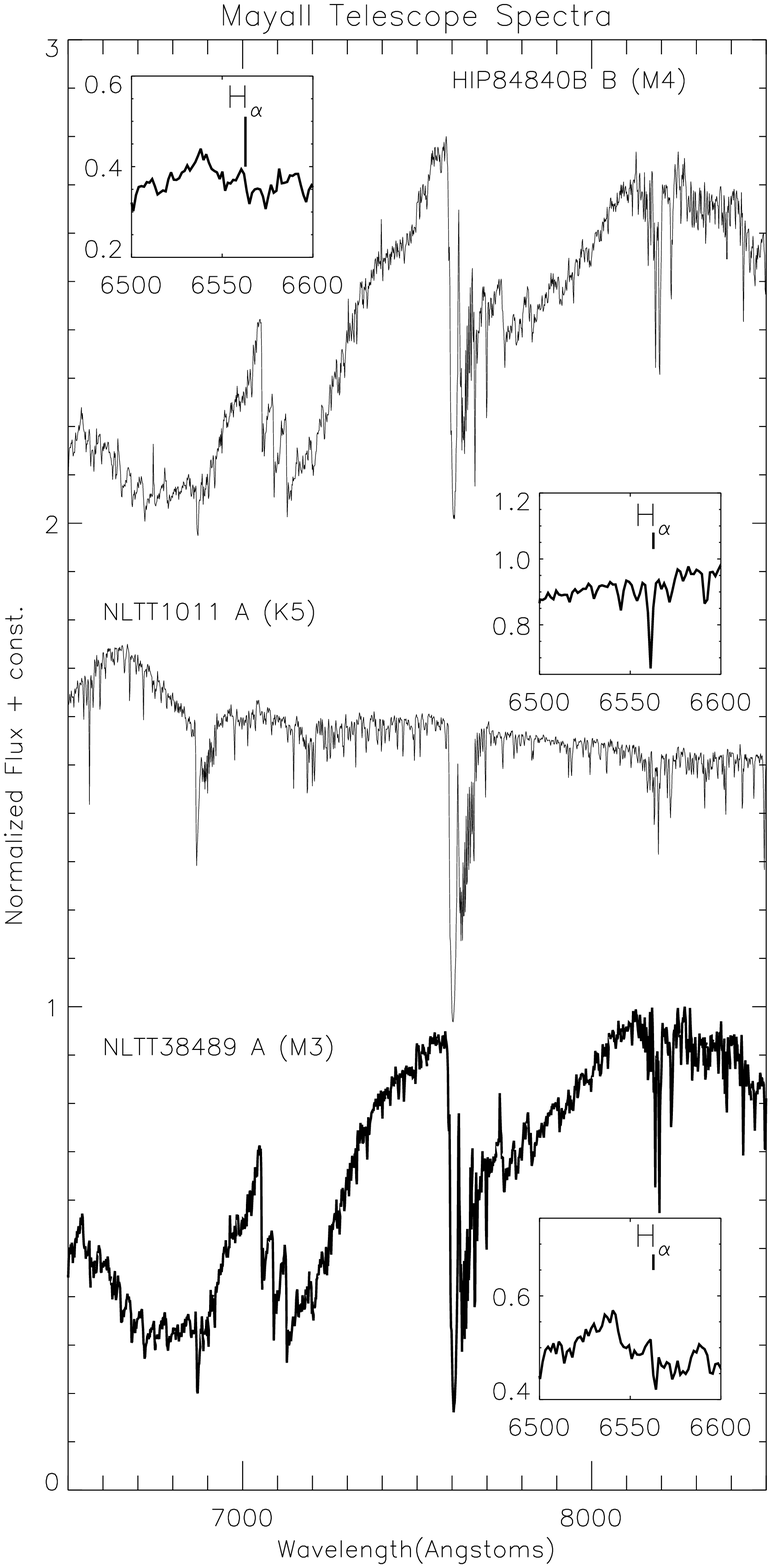}
\end{tabular}
\caption{\label{optical_primaries} Optical spectroscopy of our primary stars  and one companion. Left: spectra taken with SNIFS on the University of Hawaii 88-inch telescope on Mauna Kea. Right: spectra taken with the Ritchey-Chretien Spectrograph on the Mayall 4m telescope on Kitt Peak. The spectra for GD~280~A and LSPM~J2153+1157~A were both noisy and so have been gaussian smoothed to R=300 to make their spectral features clearer. Note the feature in the SNIFS spectra at 5200\AA is due to the boundary between the SNIFS red and blue channels.  Note the  7490-7700\AA~region is strongly affected by tellurics.}

\end{center}
\end{figure}

\begin{figure}[htbp]
\begin{center}
\epsscale{0.7}
\plotone{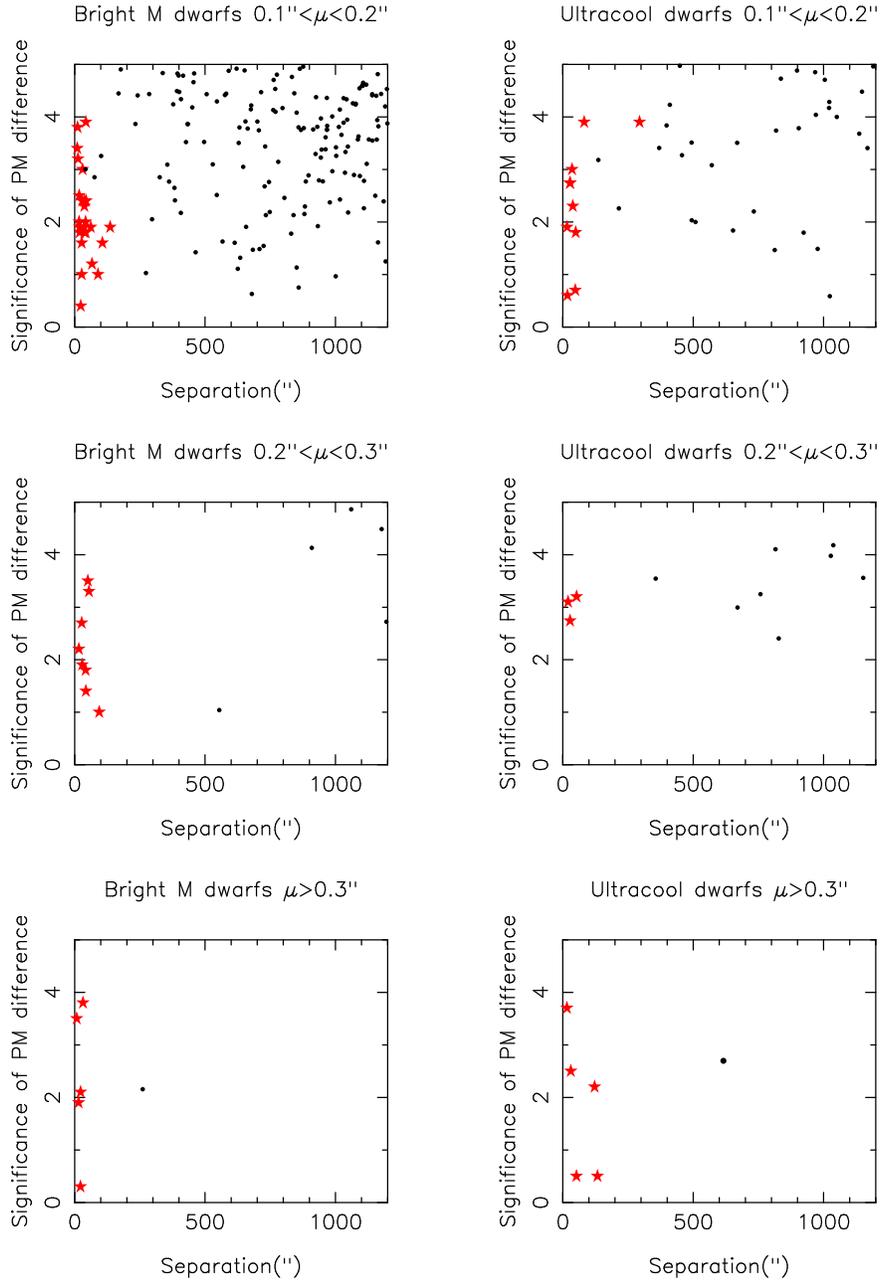}
\caption{\label{HIP_shifted} Our bright M (left column) and ultracool (right column) dwarf companions to {\it Hipparcos} stars (red stars) compared to a population of coincident objects (black dots). The coincident objects were generated in a method similar to that of \cite{Lepine2007} by offsetting the positions of primary stars in our input file and then searching for companions around these positions. The significance of proper motion difference is the quadrature sum of the proper motion difference in each axis divided by the total proper motion error on that axis (see Equation~\ref{sigmaequation}). Both of our samples lie in areas of the plot that are sparsely populated by coincident pairings.}

\end{center}
\end{figure}

\begin{figure}[htbp]
\begin{center}
\epsscale{0.35}
\plotone{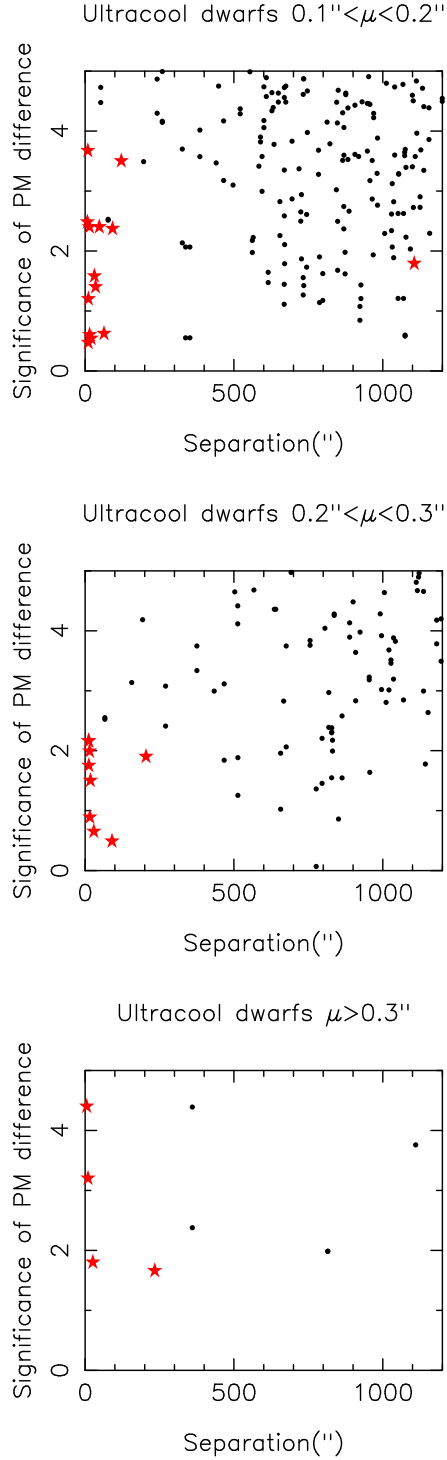}
\caption{\label{others_shifted} Our ultracool dwarf companions to faint non-HIP primaries (denoted by red stars) compared to a coincident population of objects (black dots). See Figure~\ref{HIP_shifted} caption for more details on the process. It appears one object, the apparent companion NLTT~35593 ($\mu$=0.19\,\arcsec/yr), lies in a region inhabited by many coincident pairings. Hence, despite the pair's similar photometric distances, we consider this to be an unlikely companion, i.e. one which should be confirmed through other means.}
\end{center}
\end{figure}

\begin{figure}[htbp]
\begin{center}
\epsscale{0.7}
\plotone{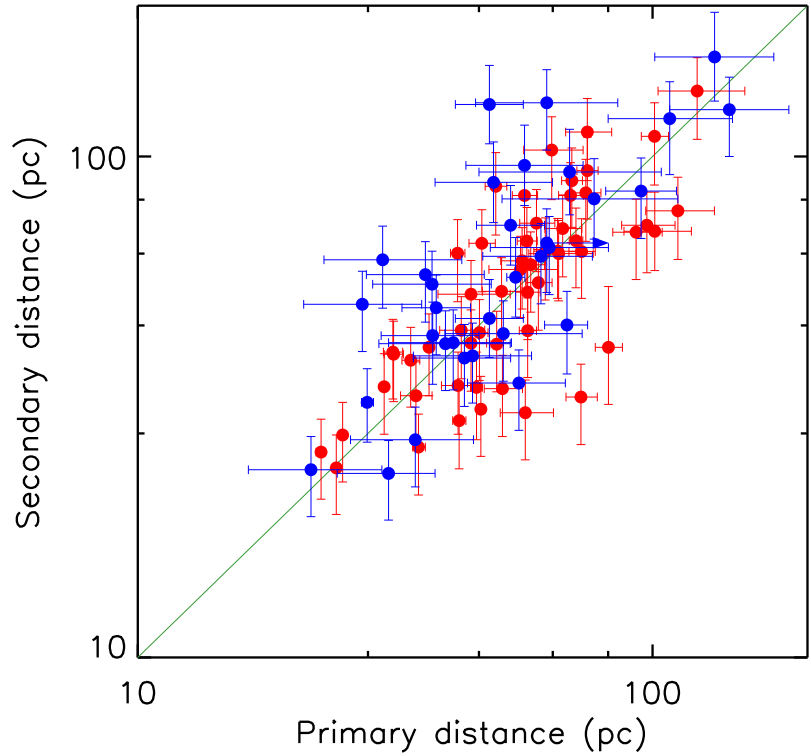}
\caption{\label{d_phot_plot} A plot of the photometric distances to our
  secondaries compared to the distances to the primaries. Points plotted in
  red have trigonometric parallaxes for their primaries, those in blue have
  photometric distance estimates. Note one of our objects (HD~253662) is known
  subgiant and hence our quoted photometric distance is a lower limit (denoted
  by an arrow).}
\end{center}
\end{figure}

\begin{figure}[htbp]
\begin{center}
\epsscale{1.0}
\plotone{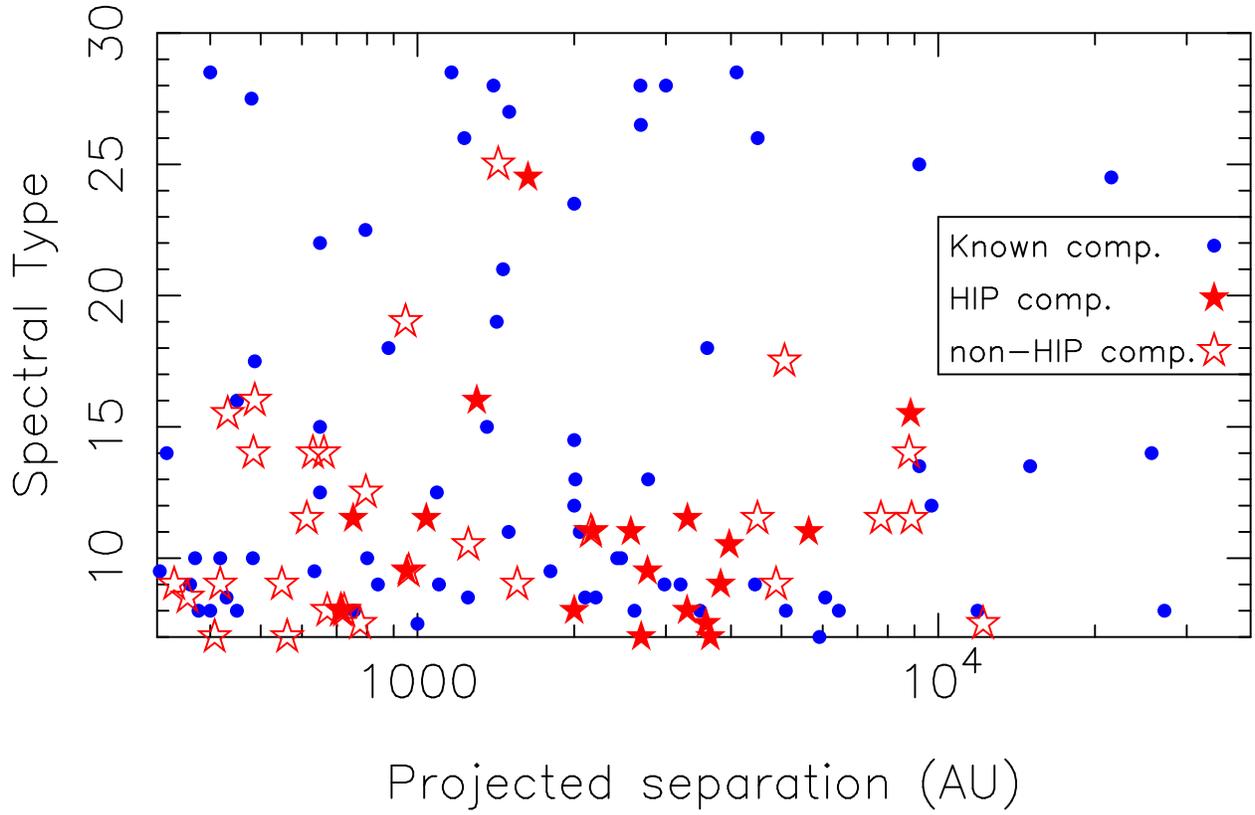}
\caption{\label{all_binaries} The field ultracool binary population (secondary
 component M7 or later). The y-axis shows the ultracool companion spectral
 type which is 0 at M0, 10 at L0 and 20 at T0. The blue dots represent previously identified objects, solid red stars are companions from
 our {\it Hipparcos} search and open red stars are our companions from other
 sources. Our two T dwarf discoveries from \cite{Deacon2012} (a {\it Hipparcos}
 companion) and \cite{Deacon2012a} (a serendipitous companion discovery) are also plotted here.}

\end{center}
\end{figure}

\begin{figure}[htbp]
\begin{center}
\epsscale{0.8}
\plotone{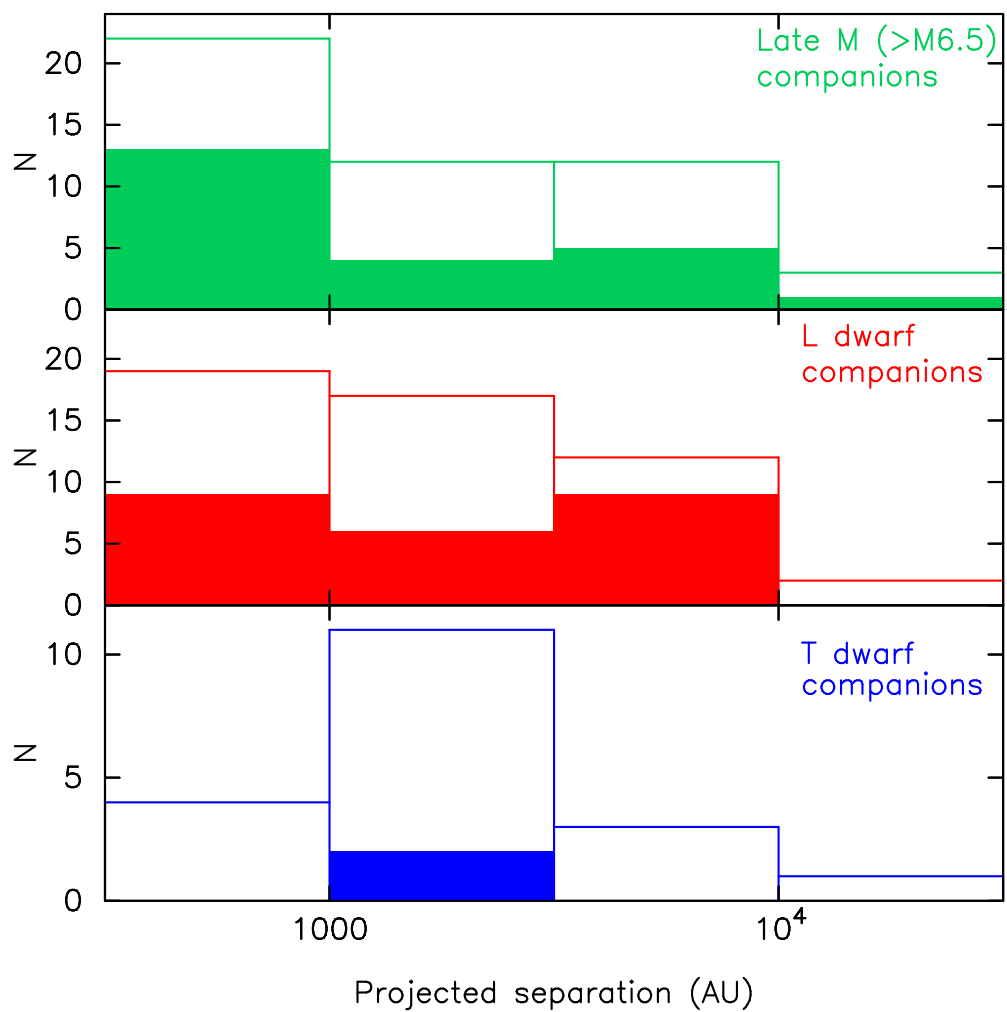}
\caption{\label{binaries_hist} Histograms showing the projected separations of
 the wide ($>$100~AU) companion population. The top panel shows M7, M8 and M9 dwarfs, the middle panel L dwarfs, and the lower panel T dwarfs. For each spectral bin, the open histogram is the total population and the solid histogram is the contribution from our PS1-based efforts (this paper, \citealt{Deacon2012} and \citealt{Deacon2012a}).}

\end{center}
\end{figure}
\begin{figure}[htbp]
\begin{center}
\epsscale{1.0}
\plotone{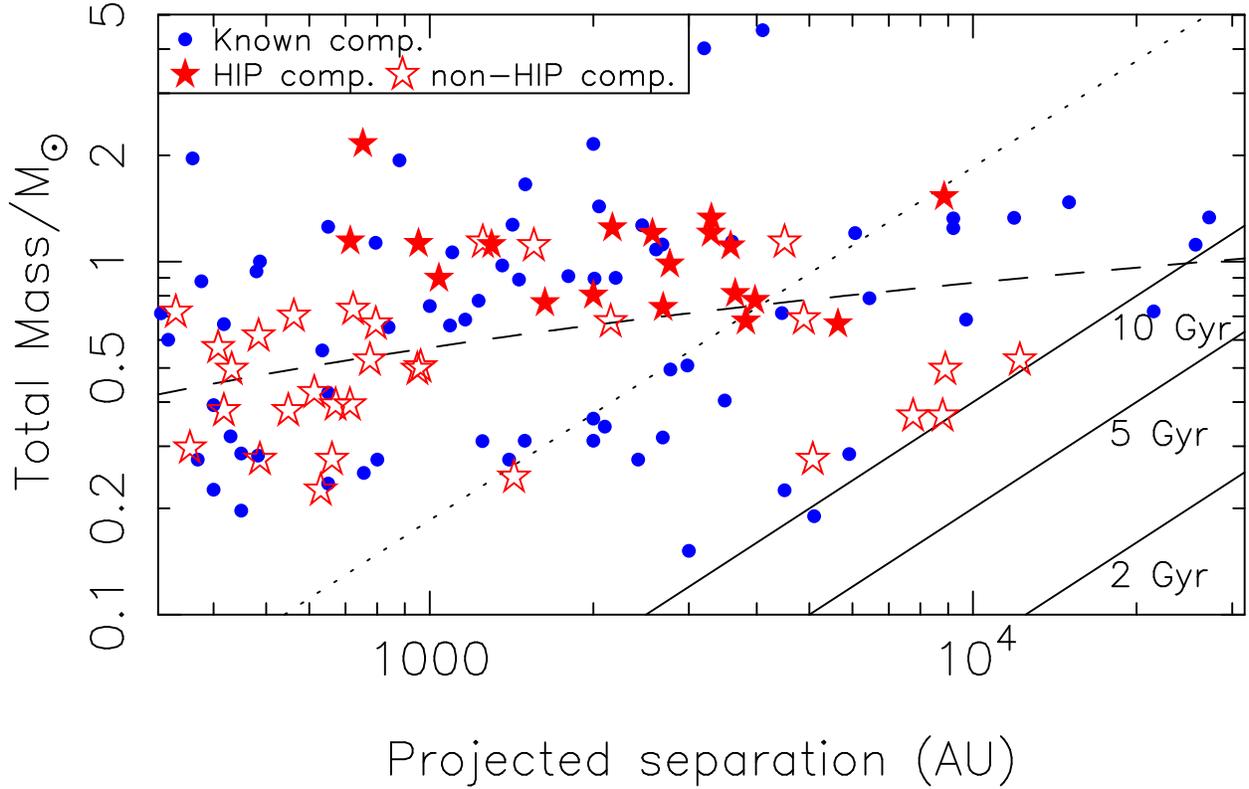}
\caption{\label{binaries_mtot} The total mass vs. separation for binary systems with at least one ultracool dwarf component from
 the literature and from our discoveries. The plot symbols are the same as
 Figure~\ref{all_binaries}. In cases where we did not have an estimated mass
 for a substellar companions, we used a mass of
 0.075\,M$_{\odot}$. Hence these are upper limits on the total mass. All other
 masses are derived from the literature or from the spectral type to mass
 relation from \cite{Kraus2007}. The dotted line represents the approximate
 maximum separation (equivalent to $v_{esc}=$0.57~km/s) suggested by
 \cite{Close2003} while the dashed line is the suggested log-normal maximum
 separation suggested by \cite{Reid2001a}. The three solid lines are the
 typical separations beyond which a binary is expected to be broken up by
 interactions in the Galactic disk over the course of 2, 5 and 10~Gyr \citep{Dhital2010}.}
\end{center}
\end{figure}

\begin{figure}[htbp]
\begin{center}
\epsscale{0.7}
\plotone{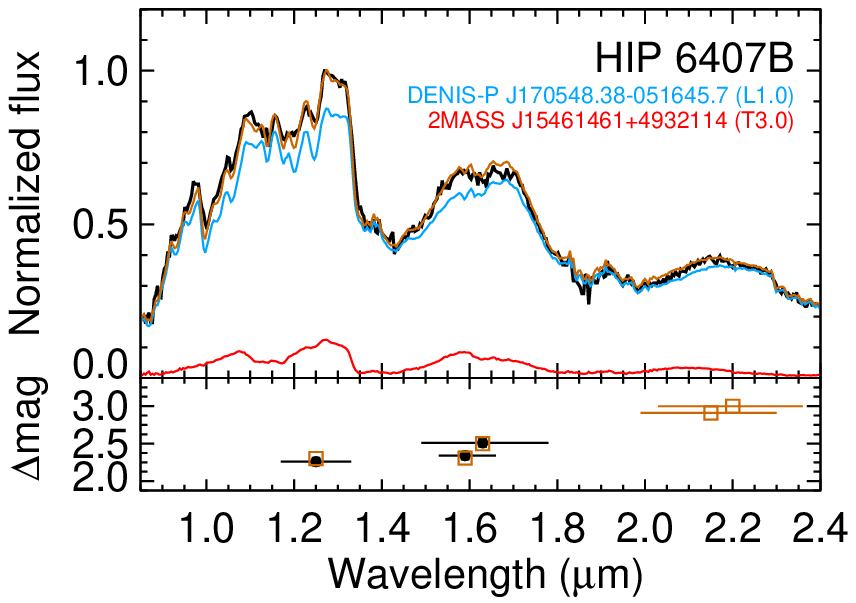}
\plotone{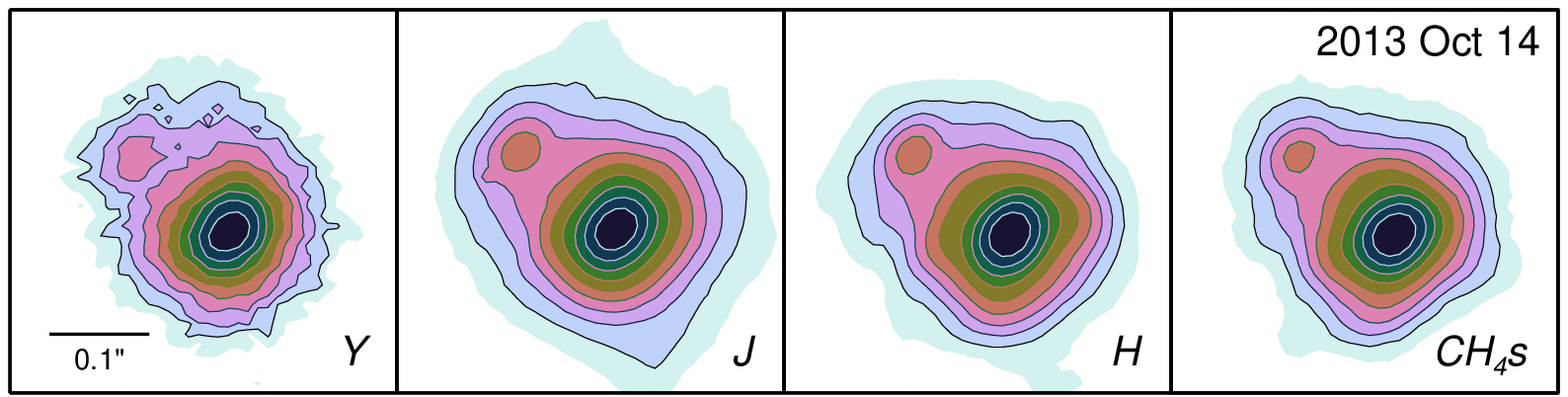}
\caption{\label{HIP_6407B} Upper panel: the spectral decomposition of HIP~6407B. The best match was the combination of the L1 dwarf DENIS-P~J170548.3$-$051645 \citep{Burgasser2010,Allers2013} and the T3 dwarf SDSS~J120602.51+281328.7 \citep{Chiu2006,Burgasser2010}. Middle panel: the resulting flux ratios between the two objects. Lower panel: Keck LGS AO images from which we derive astrometry and flux ratios with contours drawn at logarithmic intervals. Images have
been rotated such that north is up.}
\end{center}
\end{figure}

\clearpage

\begin{deluxetable}{llrr}
\tablecolumns{4}
\tablecaption{\label{surveys} Summary information on the sources for our input primary list.}
\tablehead{\colhead{Survey}&\colhead{Reference}&\colhead{$n_{total}$}&{$n_{passed}$\tablenotemark{a}}}
\startdata
{\it Hipparcos}&\cite{vanLeeuwen2007}&117955&9466\\
Lepine \& Gaidos bright M~dwarfs&\cite{Lepine2011}&8889&5432\\
LSPM North&\cite{Lepine2005}&61977&60793\\
rNLTT&\cite{Salim2003}&36085&31696\\
Total\tablenotemark{b}&&&79593\\
\tablenotetext{a}{The number of objects which passed our Galactic latitude, total proper motion and Declination cuts and (for {\it Hipparcos} stars) our parallax.}
\tablenotetext{b}{Total number excluding duplicate appearances of the same object in different catalogs.}
\enddata

\end{deluxetable}

\begin{deluxetable}{lccccccc}
\tablecaption{Keck LGS AO Observations of HIP~6407Bab on 2013 October 13 UT \label{LGStable}}
\tablewidth{0pt}
\tablehead{
\colhead{Filter} &
\colhead{$\rho$} &
\colhead{P.A.} &
\colhead{$\Delta{m}$} \\
\colhead{} &
\colhead{(mas)} &
\colhead{($^{\circ}$)} &
\colhead{(mag)} }
\startdata
$Y_{\rm NIRC2}$ & $129.7\pm2.7$ & $48.4\pm1.2$ & $2.13\pm0.06$ \\
$J_{\rm MKO}$  & $125.1\pm1.6$ & $51.1\pm1.4$ & $2.26\pm0.05$ \\
$H_{\rm MKO}$  & $125.9\pm1.0$ & $51.3\pm0.4$ & $2.51\pm0.05$ \\
$CH_4s$     & $122.4\pm2.9$ & $50.3\pm1.3$ & $2.34\pm0.05$ \\
weighted mean  & $125.8\pm0.8$ & $51.0\pm0.4$ & \nodata    \\
\enddata

\end{deluxetable}

\clearpage

\begin{landscape}
\begin{deluxetable}{lccccccccccc}
\tablecolumns{8}
\tablewidth{0pc}
\tabletypesize{\tiny}
\tablecaption{\label{all_info} Summary information on our wide companions}
\tablehead{
\colhead{Object}&
\colhead{WDS}&
\colhead{Position}&
\colhead{SpT}&
\colhead{SpT}&
\colhead{$y_{P1}$}&
\colhead{$J_{2MASS}$}&
\colhead{$d_{primary}$}&
\colhead{separation}&
\colhead{separation}&
\colhead{Metallicity\tablenotemark{k}}&
\colhead{Age}
\\
&\colhead{Designation}
&\colhead{(Eq.=J2000 Ep.=2010.0)}
&
\colhead{primary}&
\colhead{secondary}&
\colhead{mag}&
\colhead{mag}&
\colhead{pc}&
\colhead{$\arcsec$}&
\colhead{AU}&
\colhead{dex}&
\colhead{Gyr}
}
  
\startdata
\multicolumn{7}{c}{HIP search companions}\\
\hline
HIP 2397 B&\ldots&00 30 24.94 +22 44 47.1&K5&L0.5&16.606$\pm$0.008&14.586$\pm$0.038&[33.9$_{-1.3}^{+1.4}$]&117.1&3970&\ldots&0.5--$\sim$10\tablenotemark{y}\\
HIP 6217 C&\ldots&01 19 44.74 +00 06 18.5&K0&M9.5&16.859$\pm$0.009&15.441$\pm$0.060&[$70.9_{-11.7}^{+14.0}$]&27.4&2767&\ldots&0.4--$\sim$10\tablenotemark{y1}\\
HIP 6407 B&\ldots&01 22 16.99 +03 31 22.0&G5&L1+T3\tablenotemark{l}&17.527$\pm$0.027&15.471$\pm$0.046&[57.2$_{-3.0}^{+3.3}$]&44.9&2570&\ldots&0.5--$\sim$10\tablenotemark{y}\\
HIP 9269 B\tablenotemark{\ddag}&\ldots&01 59 10.98 +33 12 27.8&G5&L6&18.266$\pm$0.031&15.956$\pm$0.088&[25.0$_{-0.4}^{+0.3}$]&52.1&1300&0.15&2.2--10.2\tablenotemark{i}\\
HIP 10033 B&\ldots&02 09 10.58 $-$02 19 19.0&F8&M5&15.14$\pm$0.003&13.761$\pm$0.027&[69.7$_{-3.1}^{+3.4}$]&42.9&2990&$-$0.08&2.7--3.7\tablenotemark{j}\\
HIP 11161 B&\ldots&02 23 36.59 +52 40 05.9&F5&L1.5&18.785$\pm$0.067&16.636$\pm$0.161&[69.3$_{-3.6}^{+4.1}$]&47.7&3300&$-$0.23&0.5--2.4\tablenotemark{j}\\
HIP 13589 B\tablenotemark{*}&\ldots&02 55 07.10 +21 36 21.1&G0&M7.5&17.098$\pm$0.009&14.928$\pm$0.034&[72.8$_{-4.0}^{+4.6}$]&49.2&3580&\ldots&0.4--$\sim$10\tablenotemark{y}\\
\enddata
\tablewidth{20pc}
\vspace{-0.7cm}
\normalsize
\end{deluxetable}
\clearpage 
\end{landscape}

\begin{deluxetable}{lclccccc}
\tablecolumns{7}
\tablewidth{0pc}
\tabletypesize{\tiny}
\tablecaption{\label{HIPsecondaries} PS1 data of our wide companions.}
\tablehead{
\colhead{Object}&
\colhead{Position}&
\colhead{SpT}&
\colhead{$g_{P1}$}&
\colhead{$r_{P1}$}&
\colhead{$i_{P1}$}&
\colhead{$z_{P1}$}&
\colhead{$y_{P1}$}
\\
&\colhead{(Eq.=J2000 Ep.=2010.0)}
&
&
\colhead{mag}&
\colhead{mag}&
\colhead{mag}&
\colhead{mag}&
\colhead{mag}
}
  
\startdata
\multicolumn{7}{c}{HIP search companions}\\
\hline
HIP 2397 B&00 30 24.94 +22 44 47.1&L0.5&\ldots&\ldots&19.085$\pm$0.013&17.49$\pm$0.022&16.606$\pm$0.008\\
HIP 6217 C&01 19 44.74 +00 06 18.5&M9.5&21.656$\pm$0.11&21.076$\pm$0.071&18.6$\pm$0.019&17.444$\pm$0.007&16.859$\pm$0.009\\
HIP 6407 B&01 22 16.99 +03 31 22.0&L1+T3\tablenotemark{d}&\ldots&\ldots&20.094$\pm$0.042&18.534$\pm$0.015&17.527$\pm$0.027\\
HIP 9269 B\tablenotemark{\ddag}&01 59 10.98 +33 12 27.8&L6&\ldots&\ldots&\ldots&\ldots&18.266$\pm$0.031\\
HIP 10033 B&02 09 10.58 $-$02 19 19.0&M5&19.709$\pm$0.017&18.297$\pm$0.007&16.456$\pm$0.003&15.54$\pm$0.003&15.14$\pm$0.003\\
HIP 11161 B&02 23 36.59 +52 40 05.9&L1.5&\ldots&\ldots&21.24$\pm$0.095&19.475$\pm$0.051&18.785$\pm$0.067\\
HIP 13589 B\tablenotemark{*}&02 55 07.10 +21 36 21.1&M7.5&\ldots&\ldots&\ldots&18.003$\pm$0.012&17.098$\pm$0.009\\
\enddata
\tablewidth{20pc}
\vspace{-0.7cm}
\normalsize
\end{deluxetable} 
\begin{deluxetable}{lccccc}
\tablecolumns{5}
\tablewidth{0pc}
\tabletypesize{\tiny}
\tablecaption{\label{HIPsecondaries_ir} The IR data of our secondaries}
\tablehead{
\colhead{Object}&
\colhead{$Y$}&
\colhead{$J$}&
\colhead{$H$}&
\colhead{$K/K_{s}$}&
Source
\\
&
\colhead{mag}&
\colhead{mag}&
\colhead{mag}&
\colhead{mag}&
}
  
\startdata

\multicolumn{6}{c}{HIP search companions}\\
\hline
HIP 2397 B&\ldots&14.586$\pm$0.038\tablenotemark{*}&13.975$\pm$0.048\tablenotemark{*}&$>$13.181\tablenotemark{*}&2MASS\\
&\ldots&14.552$\pm$0.012&13.912$\pm$0.011&\ldots&UKIRT\\

HIP 6217 B&&15.441$\pm$0.060\tablenotemark{*}&14.731$\pm$0.081\tablenotemark{*}&14.423$\pm$0.079\tablenotemark{*}&2MASS\\
&\ldots&\ldots&\ldots&14.461$\pm$0.009&UKIDSS\\
HIP 6407 B&\ldots&15.471$\pm$0.046&14.651$\pm$0.061&14.402$\pm$0.08&2MASS\\
&16.442$\pm$0.008&15.345$\pm$0.005&14.738$\pm$0.006&14.203$\pm$0.006&UKIDSS\\
HIP 9269 B&\ldots&15.956$\pm$0.088\tablenotemark{*}&14.99$\pm$0.096\tablenotemark{*}&14.317$\pm$0.085\tablenotemark{*}&2MASS\\
\enddata
\tablewidth{20pc}
\vspace{-0.7cm}
\normalsize
\end{deluxetable} 

\begin{deluxetable}{lccccc}
\tablecolumns{5}
\tablewidth{0pc}
\tabletypesize{\tiny}
\tablecaption{\label{HIPsecondaries_wise} The WISE data of our companions.}
\tablehead{
\colhead{Object}&
\colhead{$W1$}&
\colhead{$W2$}&
\colhead{$W3$}&
\colhead{$W4$}&
\colhead{Note}
\\
&
\colhead{mag}&
\colhead{mag}&
\colhead{mag}&
\colhead{mag}&\\
}
  
\startdata
\multicolumn{6}{c}{HIP search companions}\\
\hline
HIP 2397 B&13.077$\pm$0.026&12.818$\pm$0.027&12.167&8.924$\pm$0.409&\\

HIP 6217 B&14.154$\pm$0.03\tablenotemark{*}&13.836$\pm$0.042\tablenotemark{*}&$>$12.790&$>$8.943&\\
HIP 6407 B&13.91$\pm$0.027\tablenotemark{*}&13.584$\pm$0.035\tablenotemark{*}&$>$12.375&$>$9.305&\\
HIP 9269 B&13.486$\pm$0.026\tablenotemark{*}&13.265$\pm$0.033\tablenotemark{*}&$>$12.573\tablenotemark{*}&9.119$\pm$0.473&\\
HIP 10033 B&12.597$\pm$0.023\tablenotemark{*}&12.33$\pm$0.023\tablenotemark{*}&12.578$\pm$0.337\tablenotemark{*}&$>$9.229&\\
HIP 11161 B&15.173$\pm$0.043\tablenotemark{*}&15.274$\pm$0.111\tablenotemark{*}&$>$12.738\tablenotemark{*}&$>$9.179&\\
HIP 13589 B&13.572$\pm$0.026\tablenotemark{*}&13.295$\pm$0.035\tablenotemark{*}&$>$12.558\tablenotemark{*}&$>$9.002&\\
\enddata
\tablewidth{20pc}
\vspace{-0.7cm}
\normalsize
\end{deluxetable} 

\begin{deluxetable}{lcccccccc}
\tablecolumns{9}
\tablewidth{0pc}
\tabletypesize{\tiny}
\tablecaption{\label{HIPsystems} Astrometric data on our data of our multiple systems.}
\tablehead{
\colhead{Object}&
\colhead{d}&
\colhead{$\mu_{\alpha}\cos\delta$}&
\colhead{$\mu_{\delta}$}&
\colhead{d$\mu_{\alpha}\cos\delta$}&
\colhead{d$\mu_{\delta}$}&
\colhead{d$\mu$}&
\colhead{Projected}&
\colhead{Projected}\\
&
&
&
&
&
&
&
\colhead{separation}&
\colhead{separation}\\
&
\colhead{pc}&
\colhead{$\arcsec$/yr}&
\colhead{$\arcsec$/yr}&
\colhead{$n_{\sigma}$}&
\colhead{$n_{\sigma}$}&
\colhead{$n_{\sigma}$}&
\colhead{$\arcsec$}&
\colhead{AU}
}
  
\tablecomments{$n_{\sigma}$ is the number of standard deviations the proper motions differ by. [\ldots] denotes distance from trigonometric parallax, all other distances are photometric.} 
\startdata
HIP 2397&[33.9$_{-1.3}^{+1.4}$]&0.186$\pm$0.001&$-$0.185$\pm$0.001&3.2&2.1&3.8&117.1&3970\\
HIP 2397B&39.2$^{+7.6}_{-6.4}$&0.2$\pm$0.004&$-$0.176$\pm$0.004\\
\hline	
HIP 6217&[101$^{+18}_{-13}$]&0.119$\pm$0.001&0.030$\pm$0.001&1.94&1.94&2.74&27.4&2767\\
HIP 6217 B&70.9$^{+14.0}_{-11.7}$&0.111$\pm$0.004&0.022$\pm$0.004\\
\hline
HIP 6407&[57.2$_{-3.0}^{+3.3}$]&0.046$\pm$0.001&$-$0.159$\pm$0.001&2.1&2.7&3.4&44.9&2570\\
HIP 6407B&53.6$^{+10.5}_{-8.8}$&0.03$\pm$0.008&$-$0.138$\pm$0.008\\
\hline
HIP 9269&[25.0$_{-0.4}^{+0.3}$]&0.244$\pm$0.001&$-$0.352$\pm$0.001&0.1&0.5&0.5&52.1&1300\\
HIP 9269B&27.8$^{+5.4}_{-4.5}$&0.243$\pm$0.011&$-$0.347$\pm$0.011\\
\enddata
\tablewidth{20pc}
\vspace{-0.7cm}
\normalsize
\end{deluxetable} 

\begin{deluxetable}{lcccccccc}
\tablecolumns{9}
\tablewidth{0pc}
\tabletypesize{\tiny}
\tablecaption{\label{othersystems} The data of our non-{\it Hipparcos} and serendipitous pairs}
\tablehead{
\colhead{Object}&
\colhead{distance}&
\colhead{$\mu_{\alpha}\cos\delta$}&
\colhead{$\mu_{\delta}$}&
\colhead{d$\mu_{\alpha}\cos\delta$}&
\colhead{d$\mu_{\delta}$}&
\colhead{d$\mu$}&
\colhead{Projected}&
\colhead{Projected}\\
&
&
&
&
&
&
&
\colhead{separation}&
\colhead{separation}\\
&
\colhead{pc}&
\colhead{$\arcsec$/yr}&
\colhead{$\arcsec$/yr}&
\colhead{$n_{\sigma}$}&
\colhead{$n_{\sigma}$}&
\colhead{$n_{\sigma}$}&
\colhead{$\arcsec$}&
\colhead{AU}
}
  
\tablecomments{$n_{\sigma}$ is the number of standard deviations the proper motions differ by. [\ldots] Denotes distance from trigonometric parallax, all other distances are photometric.} 
\startdata
\multicolumn{9}{c}{Other Companion Discoveries}\\
\hline
NLTT 1011&68.2$_{-6.5}^{+6.6}$\tablenotemark{a}&-0.076$\pm$0.008\tablenotemark{f}&-0.192$\pm$0.008&0.08&0.43&0.44&58.5&3990\\
NLTT 1011B&46.1$^{+9.3}_{-7.7}$\tablenotemark{b}&-0.077$\pm$0.007&-0.197$\pm$0.007\\ 
\hline
GD 280&77$^{+35}_{-26}$&0.171$\pm$0.008&$-$0.059$\pm$0.008\tablenotemark{f}&0.32&0.53&0.62&63.4&4882\\
GD 280 B&82.3$^{+16.7}_{-13.9}$&0.174$\pm$0.005&$-$0.054$\pm$0.005\\
\hline
NLTT 8245&53.0$^{+8.7}_{-7.5}$&$-$0.096$\pm$0.008&$-$0.164$\pm$0.008\tablenotemark{f}&0.42&0.21&0.47&10.6&562\\
NLTT 8245 B&72.9$^{+14.6}_{-12.2}$&$-$0.092$\pm$0.005&$-$0.162$\pm$0.005\\
\hline
LSPM J0241+2553&69$^{+35}_{-23}$&$-$0.032$\pm$0.008&$-$0.158$\pm$0.008\tablenotemark{f}&1.41&0.70&1.58&31.2&2153\\
\enddata
\tablewidth{20pc}
\vspace{-0.7cm}
\normalsize
\end{deluxetable}

\begin{deluxetable}{lccccccccl}
\tablecolumns{7}
\tablewidth{0pc}
\tabletypesize{\tiny}
\tablecaption{\label{SpeXlog} Log of spectroscopic observations.}
\tablehead{
\colhead{Object}&
\colhead{Date (UT)}&
\colhead{$t_{int}$}&
\colhead{Conditions}&
\colhead{Seeing}&
\colhead{Slit}&
\colhead{Mode}&
\colhead{Resolution}&
\colhead{Airmass}&
\colhead{Standard}
\\
&
&
\colhead{s}&
&
\colhead{\arcsec}&
\colhead{\arcsec}&
&
$\lambda / \Delta \lambda$&
&\\
}
  
\startdata
\multicolumn{10}{c}{HIP search companions}\\
\hline
HIP 2397 B&2012-07-05&360&cloudy&1.5&0.8$\times$15&prism&75&1.12&HD 222749\\
HIP 6217 B&2013-12-11&360&cloudy&1.3&0.5$\times$15&prism&120&1.30&HD 13936\\
HIP 6407 B&2012-07-06&480&clear&1.3&0.3$\times$15&prism&200&1.37&HD 219545\\
HIP 9269 B&2011-12-02&1440&thin clouds&0.4&0.8$\times$15&prism&75&1.156&HD 23452\\
HIP 10033 B&2012-10-25&720&cloudy&0.5&0.8$\times$15&prism&75&1.14&HD 13936\\
HIP 11161 B&2012-10-28&720&thin clouds&0.8&0.8$\times$15&prism&75&1.202&HD 21038\\
HIP 13589 B&2012-07-08&480&clear&$>$1&0.8$\times$15&prism&75&1.44&HD 19600\\
\enddata
\tablewidth{20pc}
\vspace{-0.7cm}
\normalsize
\end{deluxetable}

\begin{deluxetable}{llclclcll}
\tablecolumns{9}
\tablewidth{0pc}
\tabletypesize{\tiny}
\tablecaption{\label{Lspectratab} Spectroscopic classification of our L dwarf companions.}
\tablehead{
\colhead{Object}&
\colhead{SpT(Visual)}&
\colhead{H$_2O$\_J}&
\colhead{SpT(H$_2O$\_J)}&
\colhead{H$_2O$\_H}&
\colhead{SpT(H$_2O$\_H)}&
\colhead{CH$_4$\_K}&
\colhead{SpT(CH$_4$\_K)}&
\colhead{SpT(Final)}
\\
&
&
&
&
&
&
&
&\colhead{Type}\\
}
  
\startdata
\multicolumn{9}{c}{HIP search companions}\\
\hline
HIP 6407 B&L0&0.831&L3.3&0.776&L3.6&1.002&L4.3&L1+T5\tablenotemark{c}\\
HIP 2397 B&L0&0.935&L0.6&0.846&L0.8&1.053&L2.3&L0.5\\
HIP 26653 B&L1.5&0.900&L1.4&0.811&L2.1&1.048&L2.5&L1.5\\
HIP 59933 B&L1.5&0.922&L0.9&0.838&L1.1&1.042&L2.8&L1\\
HIP 63506 C&L1\tablenotemark{a}&0.909&L1.2&0.82&L1.8&1.025&L3.5&L1\\
HIP 85365 B&L5&0.752&L5.7&0.701&L6.5&0.965&L5.6&L5.5\\
HIP 9269 B&L6&0.752&L5.7&0.781&L3.3&0.978&L5.2&L6\\
\enddata
\tablewidth{20pc}
\vspace{-0.7cm}
\normalsize
\end{deluxetable}

\begin{deluxetable}{llcccccccccc}
\tablecolumns{12}
\tablewidth{0pc}
\tabletypesize{\tiny}
\tablecaption{\label{HIPprimary_kin} Data on our {\it Hipparcos} stars.}
\tablehead{
\colhead{Object}&
\colhead{SpT}&
\colhead{RV}&
\colhead{$H_p$}&
\colhead{$B-V$}&
\colhead{[M/H]\tablenotemark{k}}&
\colhead{[Fe/H]\tablenotemark{e}}&
\colhead{Age}&
\colhead{$v_{tan}$}&
\colhead{$U$}&
\colhead{$V$}&
\colhead{$W$}\\
&
&
\colhead{km/s}&
\colhead{mag}&
\colhead{mag}&
\colhead{dex}&
\colhead{dex}&
\colhead{Gyr}&
\colhead{km/s}&
\colhead{km/s}&
\colhead{km/s}&
\colhead{km/s}
}
\startdata
HIP 2397&K5\tablenotemark{a}&\ldots&9.34&0.93&\ldots&\ldots&0.5--$\sim$10\tablenotemark{y}&42.2&\ldots&\ldots&\ldots\\

HIP 6217&K0\tablenotemark{a}&\ldots&9.116&0.754&\ldots&\ldots&0.4--$\sim$10\tablenotemark{y*}&58.5&\ldots&\ldots&\ldots\\
HIP 6407&G5\tablenotemark{a}&\ldots&8.74&0.64&\ldots&-0.084\tablenotemark{u}&0.5--$\sim$10\tablenotemark{y}&44.9&\ldots&\ldots&\ldots\\
HIP 9269&G5\tablenotemark{a}&$-$35.3$\pm$0.1\tablenotemark{d}&7.29&0.77&0.15&\ldots&2.2--10.2\tablenotemark{k}&50.7&$-$10.9&15.8&$-$57.5\\
HIP 10033&F8\tablenotemark{a}&$-$12.9\tablenotemark{f}&7.24&0.5&\ldots&$-$0.08&2.7--3.7\tablenotemark{e}&34.3&$-$35.8&6.4&$-$4.4\\
HIP 11161&F5\tablenotemark{a}&$-$41.7\tablenotemark{f}&8.03&0.42&\ldots&$-$0.23&0.5--2.4\tablenotemark{e}&33.2&$-$45.1&$-$17.8&$-$22.2\\
HIP 13589&F8\tablenotemark{a}&\ldots&8.8&0.66&\ldots&\ldots&0.4--$\sim$10\tablenotemark{y}&61.0&\ldots&\ldots&\ldots\\
HIP 26653&G5\tablenotemark{q}&12.51$\pm$0.45\tablenotemark{r}&7.987&0.764&&$-$0.07&1.1--9.3\tablenotemark{e}\tablenotemark{*}&18.9&19.0&$-$9.1&$-$8.3\\
\enddata
\tablewidth{20pc}
\end{deluxetable}

\begin{deluxetable}{lccccccccc}
\tablecolumns{8}
\tablewidth{0pc}
\tabletypesize{\tiny}
\tablecaption{\label{othersprimary} Data on our other primary stars.}
\tablehead{
\colhead{Object}&
\colhead{SpT}&
\colhead{V}&
\colhead{V-J}&
\colhead{distance}&
\colhead{$\mu_{\alpha}\cos\delta$}&
\colhead{$\mu_{\delta}$}&
\colhead{Age}&
\colhead{[M/H]}&
\colhead{$v_{tan}$}\\
&
&
\colhead{mag.}&
\colhead{mag.}&
\colhead{pc.}&
\colhead{as/yr}&
\colhead{as/yr}&
\colhead{Gyr}&
\colhead{dex}&
\colhead{km/s}
}
\startdata

\multicolumn{8}{c}{Other Companion Discoveries}\\
\hline 
NLTT 1011&K5\tablenotemark{b}&11.13&2.0&68.2$_{-6.5}^{+6.6}$&$-$0.076&$-$0.192&0.3--$\sim$10\tablenotemark{i}&\ldots&66.8\\
GD 280&DA\tablenotemark{b}&16.69&0.02&77$^{+35}_{-26}$&0.169&$-$0.055&$<$10\tablenotemark{k}&\ldots&\\
NLTT 8245&M0\tablenotemark{c}&12.63&2.98&53.0$^{+8.7}_{-7.5}$&$-$0.093&$-$0.164&0.3--$\sim$10\tablenotemark{i}&\ldots&47.4\\
LSPM J0241+2553&WD\tablenotemark{l}&18.19&1.19&69$^{+35}_{-23}$&$-$0.037&$-$0.153&$<$10\tablenotemark{k}&\ldots&\\
HD 253662&G8IV\tablenotemark{d}&9.92&1.36&$>$62.3&$-$0.013&$-$0.164&0.3--$<$10\tablenotemark{k}&\ldots&\\
LSPM J0632+5053&G2\tablenotemark{b}&9.87&1.26&95.0$_{-13.9}^{+16.3}$&0.035&$-$0.154&0.2--$\sim$10\tablenotemark{g}&\ldots&71.1\\
NLTT 18587&M2\tablenotemark{c}&16.07&3.56&132$^{+40}_{-31}$&0.177&$-$0.109&0.3--$\sim$10\tablenotemark{i}&\ldots&130\\
\enddata
\tablewidth{20pc}
\end{deluxetable}

\vspace{-0.7cm}
\normalsize

\clearpage
\begin{landscape}
\begin{deluxetable}{lccrllccl}
\tablecolumns{12}
\tablewidth{0pc}
\tabletypesize{\tiny}
\tablecaption{\label{known_comp} Known companions with spectral type M7 or later which have projected separations of 100~AU or more.}
\tablehead{
\colhead{Object}&
\colhead{Position}&
\multicolumn{2}{c}{Separation}&
\colhead{SpT}&
\colhead{SpT}&
\colhead{companion mass}&
\colhead{age}&
\colhead{References}
\\
&
\colhead{(J2000)}&
\colhead{AU}&
\colhead{\arcsec}&
\colhead{companion}&
\colhead{primary}&
\colhead{M$_{\odot}$}&
\colhead{Gyr}&
}
\startdata
2MASS J1258+4013 B&12 58 37.98 +40 14 01.7&6700&63&M7&M6&0.086-0.105&1-5&1\\
GG Tau Bb&04 32 30.31 	+17 31 29.9 \tablenotemark{o}&210&1.5&M7&K7+M0.5+M5&0.038-0.05&0.01-0.02&2\\
HD 65216 B&07 53 42.55 $-$63 38 51.5\tablenotemark{p}&253&7.0&M7&G5&0.09&3-6&3\\
~~~~~~~~~~~~C&&&&L2&&0.078&&\\
$\eta$ Tel B&19 22 51.26 $-$54 25 30.7&190&4.2&M7/8&A0V&$<$0.04&0.0017-0.025\tablenotemark{a}&4\\
G121$-$42 B&12 00 32.92 +20 48 51.3\tablenotemark{q}&5916&204.0&M7&M4&$<$0.085&4.0-5.0&5\\
USco 1602$-$2401&16 02 51.17 $-$24 01 50.5&1000&7.0&M7.5&K4&0.029-0.067&0.010-0.013\tablenotemark{k}&65\\
TWA 5 B&11 31 55.4 -34 36 29&100&2.0&M8&M1.5&$\sim$0.02&0.01-0.3&6\\
\enddata
\tablewidth{20pc}
\end{deluxetable}
\end{landscape}
\end{document}